\documentclass[acmlarge,table]{acmart}

\setcitestyle{nosort}

\usepackage{amsmath}
\usepackage[T1]{fontenc}
\usepackage{multicol}
\usepackage{booktabs}
\usepackage{okexplain}
\usepackage{suffix}
\usepackage{oktodo}
\usepackage{url}
\usepackage{subfig}
\usepackage{xspace}
\usepackage{upgreek}

\newcommand{\citepos}[1]{\citeauthor{#1}'s~\citeyearpar{#1}}
\WithSuffix\newcommand\citepos*[2]{\citeauthor{#1}'s #2~\citeyearpar{#1}}
\WithSuffix\newcommand\citepos-[2]{\citeauthor{#1}#2~\citeyearpar{#1}}

\newcommand\felleisen{felleisen:on-the-expressive-power-of-programming-languages}
\newcommand\hia{kammar-lindley-oury:handlers-in-action}
\newcommand\mia{filinski:monads-in-action}

\newcommand\lang{\mathcal L}

\newcommand\cbpv{\text{\textsc{mam}}\xspace}
\newcommand\lameff{\textrm{\textsc{eff}}\xspace}
\newcommand\lammon{\textrm{\textsc{mon}}\xspace}
\newcommand\lamdel{\textrm{\textsc{del}}\xspace}
\newcommand\typed{\textrm{\textsc{typed }}}
\newcommand\proper{\typed}
\newcommand\typeable{\typed}

\newcommand\figlabel[1]{\label{fig:#1}}
\newcommand\figref[1]{Fig.~\ref{fig:#1}}
\WithSuffix\newcommand\figref*[1]{\ref{fig:#1}}
\newcommand\figmul{Fig.}
\newcommand\subfiglabel[1]{\label{subfig:#1}}
\newcommand\subfigref[2]{Fig.~\ref{fig:#1}\subfigref*{#2}}
\WithSuffix\newcommand\subfigref*[1]{(\ref{subfig:#1})}

\newcounter{subcaption}[figure]
\renewcommand\thesubcaption{\alph{subcaption}}

\newcommand\subcaptionn[2]{
  \refstepcounter{subcaption}\subfiglabel{#2}
  (\thesubcaption) #1
\smallskip}

\usepackage{oksemantics}
\usepackage{okcategory}
\usepackage{oktheorem}
\usepackage{mathpartir}

\usepackage[table]{xcolor}

\definecolor{shade}{RGB}{223,223,223}
\definecolor{unshade}{RGB}{255,255,255}

\usepackage{tcolorbox}
\newtcbox{\shadebox}{on line,arc=1pt, outer arc=2pt,
  colback=shade,colframe=shade,boxsep=0pt,
  left=1pt,right=1pt,top=2pt,bottom=2pt,
  boxrule=0pt,bottomrule=1pt,toprule=1pt}
\newcommand{\shade}[1]{
        \shadebox{\ensuremath{#1}}
}
\newtcbox{\unshadebox}{on line,arc=1pt, outer arc=2pt,
  colback=unshade,colframe=shade,boxsep=0pt,
  left=1pt,right=1pt,top=2pt,bottom=2pt,
  boxrule=0pt,bottomrule=1pt,toprule=1pt}

\newcommand\figheading[1]{{\textbf{#1}}}

\newcommand\ba{\begin{array}}
\newcommand\ea{\end{array}}

\newcommand\slab[1]{(\textrm{#1})}

\newcommand\V{V}
\newcommand\W{W}
\newcommand\M{M}
\newcommand\N{N}
\renewcommand\H{H}

\providecommand\G{}
\renewcommand\G{\Gamma}
\newcommand\g{\gamma}
\newcommand\D{\Delta}

\newcommand\ValueKind{\mathbf{Val}}
\newcommand\ComputationKind[1]{\mathbf{Comp}_{#1}}
\newcommand\EffectKind{\mathbf{Eff}}
\newcommand\ContextKind{\mathbf{Ctxt}}
\newcommand\HandlerKind{\mathbf{Hndlr}}

\newcommand\vUnitType{1}
\newcommand\vProdType[2]{{#1} \times {#2}}

\newcommand\VariantType[1]{\{#1\}}
\newcommand\VariantConstructor[2]{\inj{#1}{#2}}
\newcommand\vand{\mid}

\newcommand\SumZero{0}

\providecommand\U{}
\renewcommand\U[2]{U_{#1}#2}

\newcommand\F[1]{F#1}
\newcommand\ProdType\set
\newcommand\Fun[2]{#1\to#2}

\newcommand\cProdType[2]{{#1} \mathbin{\&} {#2}}

\newcommand\Arity[2]{\of {#2} \to {#1}}
\newcommand\Ar{\B}
\newcommand\Pa{\A}
\newcommand\HandlerType[4]{{#1} \mathbin{{}^{#2}\mathord{\Rightarrow}^{#4}} {#3}}

\newcommand\CompType[1]{#1}
\newcommand\A{A}
\newcommand\B{B}
\providecommand\C{}
\renewcommand\C{\CompType{C}}
\newcommand\Cv{\CompType{D}}
\newcommand\Gnd{G}
\newcommand\HT{R} 
\newcommand\e{E}

\newcommand\vinf[3]{;{#1}\entails{#2}\of{#3}}
\newcommand\cinf[4][\E]{;{#2}\entails_{#1}{#3}\of{#4}}
\newcommand\einf[4]{;#2\entails_{#1}#3\of#4}
\newcommand\hinf[3]{;{#1}\entails#2\of{#3}}

\newcommand\kinf[3][\Ty]{#1\entails_{\mathrm k}#2 \of #3}
\newcommand\geninf[5]{#1;#2\entails_{#3}#4\of#5}

\newcommand\betavprod{\times}
\newcommand\betaplus{+}
\newcommand\betaU{U}
\newcommand\betaF{F}
\newcommand\betafun{\mathord{\to}}
\newcommand\betacprod{\&}

\newcommand\handleOp{\mathit{op}}
\newcommand\handleReturn{\mathit{ret}}

\newcommand\reifyRule{\textit{\normalsize ret}}
\newcommand\reflectRule{\textit{\normalsize reflection}}

\newcommand\resetRule{\textit{\normalsize ret}}
\newcommand\shiftRule{\textit{\normalsize capture}}

\newcommand\reducesto\leadsto
\newcommand\reducestoupto{\reducesto_{\mathrm{cong}}}
\WithSuffix\newcommand\reducestoupto+{\reducesto^+_{\mathrm{cong}}}

\WithSuffix\newcommand\reducesto+{\leadsto^+}
\newcommand\breducesto{\reducesto_{\beta}}
\newcommand\smallstep[2]{{#1}\reducesto{#2}}
\newcommand\substitute[2]{{#1}[{#2}]}
\providecommand\for{}
\renewcommand\for[2]{{#1}/{#2}}

\newcommand\ignore[1]{}    

\newcommand\gor{{}\mathrel{\vert}{}} 
\newcommand\gdefinedby{\mathrel{:}\mathrel{:}=} 
\newcommand\entails{\vdash}

\newenvironment{syntax}{\[
  \rowcolors{100}{white}{white}
  \ba{@{}l@{~~}r@{~~}c@{~}l@{}}
}{
  \ea
  \]
}

\newcommand\syncat[1]{\mspace{-25mu}\synname{#1}}
\newcommand\synname[1]{\qquad\text{#1}}

\newenvironment{newsyntax}[1][]{
\(
  \rowcolors{100}{white}{white}
  \begin{array}[t]{#1l@{\quad\!\!}*3{l@{}}@{\,}l}
}{
\end{array}
\)
}

\newenvironment{reductions}[1][]{\(
  \rowcolors{100}{white}{white}
  \begin{array}[t]{#1@{}l@{\quad}r@{~}c@{~}l@{~~~~~}l}
}{
  \end{array}
  \)
}

\newcommand\keyword\mathbf

\newcommand\CF{\mathcal{H}}  
\newcommand\EF{\mathcal{F}}  
\newcommand\EB{\mathcal{B}}  
\newcommand\EC{\mathcal{C}}

\newcommand\Ctx{\mathcal{X}}

\newcommand\hole{[\ \ ]}

\newcommand\varstyle[1]{{\mathit #1}}
  \newcommand\x{\varstyle{x}}
  \newcommand\y{\varstyle{y}}
  
\renewcommand\k{\varstyle{k}}
  \newcommand\p{\varstyle{p}}

\newcommand\var\varstyle

\makeatletter
\def\MathparLineskip{\lineskip=.6em plus .1em minus 0em}
\makeatother

\newcommand\vunit{\texttt{()}}

\newcommand\inj[2]{\keyword{inj}_{#1}\,#2}

\newcommand\thunk[1]{\{#1\}}

\newcommand{\retK}{\keyword{return}}

\newcommand{\synpair}[2]{\texttt{(}#1,#2\texttt{)}}
\newcommand\vpair\synpair

\newcommand\Match[3][]{\keyword{case}\ #2\ \keyword{of}#1\ #3}
\newcommand\MatchCase[2]{#1\to#2}
\newcommand\Split[5][]{\Match[{#1}]{#2}{\MatchCase{\vpair{#3}{#4}}{#5}}}
\newcommand\Variant[3][]{\Match[{#1}]{#2}{\{#3\}}}
\newcommand\VariantCase[3]{\MatchCase{\inj{#1}{#2}}{#3}}

\newcommand\force[1]{{#1}!}
\usepackage{xparse}
\newcommand{\return}[2][]{
  {\ifx&#2&
      \retK_{#1}
  \else
      \retK_{#1}\ #2
  \fi}
}

\newcommand\Let[3]{#1 \gets #2;\ #3}

\newcommand\abst[2]{\lambda#1.#2}
\newcommand\apply[2]{{#1}\ {#2}}

\newcommand\Entry[2]{#1 \mapsto #2}

\newcommand\cpair[2]{\anglebrackets{{#1},{#2}}}
\newcommand\proj[2]{\keyword{prj}_{#1}\,{#2}}

\newcommand\ShiftZ[2]{\keyword{S_0}#1.#2}

\providecommand\Reset{}
\renewcommand\Reset[3]{\left<#1\middle\lvert#2.#3\right>}
\newcommand\Resetk[2]{\left<#1\middle\lvert#2\right>}

\newcommand\setindex[2]{\{#1\}_{#2}}
\newcommand\seti[2][i]{\setindex{#2}{#1}}

\newcommand\sem\scottbrackets
\newcommand\ksem[2][\tysem]{\sem{#2}_{#1}}
\newcommand\tsem[2][\tysem]{\sem{#2}_{#1}}
\newcommand\bind[1][]{{\mathrel{\gg\!\!=_{#1}}}}
\newcommand\fmap{\keyword{fmap}}

\newcommand\carrier[1]{\left\lvert{#1}\right\rvert}

\newcommand\ceq {\simeq}

\newcommand\ds\widehat
\newcommand\cps[1]{#1}
\newcommand\mds\hat

\newcommand\opstyle{\textsf}
\newcommand\op[1][op]{\opstyle{#1}}
\newcommand\operation[2]{\cps{#1}~{#2}}

\newcommand\handle[2]{\keyword{handle}\ #1\ \keyword{with}\ #2}
\newcommand\handlerEntry[4]{\Entry{#1\,{#2}\,{#3}}{#4}}
\newcommand\returnClause[2]{\Entry{\retK~{#1}}{#2}}

\newcommand\opClauseOpen{\{}
\newcommand\opClauseClose{\}}

\newcommand\handler[3]{{#3} \mathbin{\bullet}\, \returnClause{#1}{#2}}
\WithSuffix\newcommand\handler*[3]{\returnClause{#1}{#2}\mathbin{\bullet}\,\opClauseOpen #3 \opClauseClose}

\newcommand\handlerOp[2]{{#1}^{#2}}
\newcommand\handlerRet[1]{{#1}^{\retK}}

\newcommand\reifybrackets[1]{\textnormal{\texttt{[}}#1\textnormal{\texttt{]}}}

\newcommand\reflect[1]{\upmu(#1)}

\newcommand\reify[2][\Monad]{\reifybrackets{#2}^{#1}}

\newcommand\effdef[2]{\mathop{\keyword{where}}\,\{#1 ;  #2\}}
\WithSuffix\newcommand\effdef*[6][]{\effdef{#1\return #2 = #3}{#4 \bind #5 = #6}}

\newcommand\ta\alpha
\newcommand\tb\beta
\newcommand\tc\gamma

\newcommand\T{\mathrm{T}}

\newcommand\menv\Phi

\newcommand\sig{\Sigma}

\newcommand\effType[4]{#1\prec \keyword{instance\ monad}\parent{{#2}.{#3}}#4}
\WithSuffix\newcommand\effType*[5][]{#2\prec \keyword{instance\ monad}#1\parent{{#3}.{#4}}#5}

\newcommand\arity[1][]{\mathop{arity}\nolimits_{#1}}

\DeclareDocumentCommand{\monvinf}{O{\shade \e} m m m O{\Theta\ |\ }}{{{#5 #2}\entails_{\scriptsize #1}{#3}\of{#4}}}
\newcommand\minf[2]{\entails_{\mathrm m}#1:#2}
\WithSuffix\newcommand\minf*[4]{\minf{#1}{\effType{#2}{#3}{#4}{#1}}}

\usepackage{mathtools}
\newcommand{\sharedSyntax}[2]{
  &\mathrlap{\M, \N\gdefinedby #1}& &\syncat{computations} \\
    #2
}

\newcommand{\groundTypes}{
    \slab{ground values} &
   \Gnd &\gdefinedby&
                 \vUnitType
            \gor \vProdType{\Gnd_1}{\Gnd_2}
            \gor \VariantType{\VariantConstructor{\ell_1}{\Gnd_1}\vand\ldots\vand\VariantConstructor{\ell_n}{\Gnd_n}}
}

\newcommand{\sharedValueTyping}{}

\newcommand\sharedFrames{
                      \Let{x}{[~]}{\N}
         \gor         \apply{[~]}{\V}
         \gor         \proj{i}{[~]}
}

 \newcommand\cbpvSyntax{
  \begin{newsyntax}
    &\mathrlap{\V, \W \gdefinedby}&      &\syncat{values}             \\
&    & \x                & \synname{variable}   \\
    &\gor& \vunit            & \synname{unit value} \\
    &\gor& \mathrlap{\vpair{\V_1}{\V_2}}\mspace{10mu}
    & \synname{pairing}    \\
    &\gor& \inj{\ell}{\V}       & \synname{variant}\\
    &\gor& \thunk\M          & \synname{thunk}
  \end{newsyntax}
  \qquad
  \begin{newsyntax}[@{}]
  \sharedSyntax{}{
    &    & \Split[{& \synname{product}\\&&\quad}]{\V}{\x_1}{\x_2}{\M\mspace{-100mu}\hphantom{a} & \synname{matching}}\\
    &\gor& \Variant{\V}{ &\synname{variant}\\
    &    &\quad \VariantCase{\ell_1}{\x_1}{\M_1}  \mspace{-30mu}\hphantom{a}& \synname{matching}\\
    &    &\quad \smash\vdots \\&&\quad\VariantCase{\ell_n}{\x_n}{\M_n}
    }                          \mspace{-80mu}\hphantom{a}&\\
  \end{newsyntax}\begin{newsyntax}
    &\gor&  \force \V &\synname{force} \\
    &\gor& \mathrlap{\return \V} &\synname{returner}\\
    &\gor& \mathrlap{\Let{\x}{\M}{\N}} & \synname{sequencing} \\
    &\gor& \abst \x\M       & \synname{abstraction}\\
    &\gor& \apply \M\V & \synname{application} \\
    &\gor& \mathrlap{\cpair{\M_1}{\M_2}} & \synname{pairing} \\
    &\gor& \proj{i}{\M}     & \synname{projection}
}
\end{newsyntax}
}

\newcommand{\cbpvTypes}{
  \hspace{-1em}
\begin{newsyntax}
  &\mathrlap{\E    \gdefinedby}& & \syncat{effects}\\
        &           & \pure & \synname{pure effect}\\
  &\mathrlap{K      \gdefinedby}& &\syncat{kinds}\\
        &\gor& \EffectKind & \synname{effects}\\
        &\gor& \ValueKind  & \synname{values} \\
        &\gor& \mathrlap{\ComputationKind \E} & \synname{computations} \\
        &\gor& \ContextKind &  \synname {environments} \\

\end{newsyntax}\quad\begin{newsyntax}
  &\mathrlap{\A,\B\gdefinedby}&\mspace{40mu}\hphantom{a} & \syncat{value types}\\
        &    & \tyv                   & \synname{type variable} \\
        &\gor& \vUnitType             & \synname{unit}    \\
        &\gor& \mathrlap{\vProdType{\A_1}{\A_2}} & \synname{products} \\
        &\gor& \VariantType{\mathrlap{\VariantConstructor{\ell_1}{\A_1}}  & \synname{variants}            \\
        &    & \vand\ldots \vand
              \mathrlap{ \VariantConstructor{\ell_n}{\A_n}\}}\ignore}                       & \\
        &\gor& \U{\E}\C               & \synname{thunks}  \\

\end{newsyntax}\quad\begin{newsyntax}
  &\mathrlap{\C,\Cv\gdefinedby}& & \syncat{computation types} \\
        &    &  \F\A      & \synname{returners} \\
        &\gor&  \Fun \A\C & \synname{functions} \\
  &\gor&  \cProdType{\C_1}{\C_2} & \synname{products} \\
  \multicolumn{4}{l}{\text{Environments:}}\\
  &\multicolumn3{l}{\Ty  \gdefinedby{} \tyv_1, \ldots, \tyv_n}\\
  &\multicolumn3{l}{\mathrlap{\G,\D\gdefinedby \x_1\of\A_1, \ldots, \x_n\of\A_n }}

\end{newsyntax}
}

\newcommand{\cbpvOpSemantics}{
\begin{tabular}{ll}
  \begin{tabular}[t]{l}
  \figheading{Frames and contexts}\\
  \begin{newsyntax}
    \EB   &\gdefinedby{}& \sharedFrames& \syncat{basic frames}       \\
    \EF   &\gdefinedby  & \EB          & \syncat{computation frames} \\
    \EC   &\gdefinedby&
                \hole\gor\EC[\EF\hole] & \syncat{evaluation context} \\
    \CF   &\gdefinedby&
                \hole\gor\CF[\EB\hole] & \syncat{hoisting context}
  \end{newsyntax}
    \\\\
  \figheading{Reduction} $\quad \boxed{\smallstep{\M}{\M'}}$\\
  \(
    \inferrule
      {\M \breducesto \M'}
      {\EC[\M] \reducesto \EC[\M']}
  \)
    \end{tabular}
          &
  \begin{tabular}[t]{l}
  \figheading{Beta reduction} $\quad \boxed{{\M}\breducesto{\M'}}$
    \medskip\\
  \begin{reductions}
    (\betavprod) &
    \multicolumn{3}{l}{\Split{\vpair{\V_1}{\V_2}}{\x_1}{\x_2}{\M}}
      \\&&\breducesto&
        \substitute\M{\for{\V_1}{\x_1}, \for{\V_2}{\x_2}}
    \\
    (\betaplus) &
    \multicolumn{3}{l}{\Variant{\inj{\ell}{\V}}{\ldots\VariantCase{\ell}{\x}{\M}\ldots}}
      \\&&\breducesto&
    \substitute{\M}{\for\V\x}
    \\
    (\betaF) &
    \Let\x{\return \V}\M &\breducesto& \substitute\M{\for \V\x}
    \\
    (\betaU) &
    \force{\thunk\M} &\breducesto& \M
    \\
    (\betafun) &
    \apply{(\abst\x\M)}\V &\breducesto& \substitute\M{\for \V\x}
    \\
    (\betacprod) &
    \proj{i}{\pair{\M_1}{\M_2}} &\breducesto& \M_i
  \end{reductions}
  \end{tabular}
\end{tabular}
}

\newcommand{\cbpvDenSem}{
\begin{tabular}{@{}l@{~}l@{}}
\begin{tabular}[t]{l@{\,}}
\figheading{Effects}
\hspace{1.025cm}
\(
  \ksem{\pure} \definedby \triple\Id\id\id
\)
\\[1ex]
\figheading{Value types}
\quad
\(  \ksem{\tyv} \definedby \tysem(\tyv) \quad   \ksem{\vUnitType} \definedby \set{\star}
\)\\
\(\begin{array}[t]{@{}l@{}}
  \ksem{\vProdType{\A_1}{\A_2}} \definedby \ksem{\A_1}\times\ksem{\A_2}
  \
  \ksem{\U{\E}\C} \definedby \carrier{\ksem \C}
  \\
  \begin{array}[t]{@{}l@{}}
  \ksem{\VariantType{\VariantConstructor{\ell_1}{\A_1}
                    \vand\ldots\vand
                        \VariantConstructor{\ell_n}{\A_n}}} \definedby
  \\\quad\quad(\set{\ell_1}\times\ksem{\A_1})\union\cdots
  \union(\set{\ell_n}\times\ksem{\A_n})
  \end{array}
  \end{array}
\)
\end{tabular}
&
\begin{tabular}[t]{l}
\figheading{Computation types} \\
\(
\begin{array}[t]{l}
  \ksem{\F\A} \definedby F{\ksem{\A}}
  \\
  \ksem{\Fun \A\C} \definedby
  \pair{\smash{{\carrier{\ksem{\C}}}^{\ksem{A}}}}{\lambda f_s.\lambda x.c(\fmap\ (\lambda
    f.f(x))\ f_s)}
  \\
  \begin{array}[t]{@{}l@{}}
  \ksem{\cProdType{\C_1}{\C_2}} \definedby
  \\\quad
  \pair{\carrier{\ksem{\C_1}}\times\carrier{\ksem{\C_2}}}{\lambda
    c_s.\pair{c_1(\fmap\ \projection_1\ c_s)}{c_2(\fmap\
      \projection_2\ c_s)}}
  \end{array}
\end{array}
\)
\end{tabular}
\end{tabular}
}

\newcommand{\cbpvTermDenSem}{
\begin{tabular}{@{}l@{}}
  \figheading{Value terms}\qquad
  \(\def\SharedSpace{\mspace{50mu}}
    \begin{array}[t]{@{}l@{\SharedSpace}l@{\SharedSpace}l@{}}
    \tsem{\x}(\gamma) \definedby \projection_{\x}(\gamma)
    &
      \tsem{\vpair{\V_1}{\V_2}}(\gamma) \definedby
      \pair{\tsem{\V_1}(\gamma)}{\tsem{\V_2}(\gamma)}
\\
    \tsem{\vunit}(\gamma) \definedby \star
    &
    \tsem{\inj{\ell}{\V}}(\gamma) \definedby \pair {\ell}{\tsem\V(\gamma)}
    &
    \tsem{\thunk\M}(\gamma) \definedby \tsem\M(\gamma)
\end{array}\)
\\
\figheading{Computation terms} \\
\(
\begin{array}{@{}l@{\mspace{115mu}}l@{}}
  \multicolumn{2}{@{}l@{}}{
   \begin{array}{@{}l@{\quad}l@{}}
     \tsem{\Split{\V}{\x_1}{\x_2}{\M}}(\gamma) \definedby
     \tsem\M(\gamma[\x_1 \mapsto a_1, \x_2 \mapsto a_2])
   & \text{ where
            $\tsem{\V}(\gamma) = \pair{a_1}{a_2}$} \\
     \tsem{
         \Variant{\V}{
           \VariantCase{\ell_1}{\x_1}{\M_1}
           \cdots
           \VariantCase{\ell_n}{\x_n}{\M_n}}
     }
     \definedby
     \tsem{\M_i}(\gamma[\x_i \mapsto a_i])
     & \text{ where $\tsem{V}(\gamma) = \pair{\mathrlap{\ell_i}\mspace{17mu}}{a_i}$}
    \end{array}
  }
\\
  \tsem{\force\V}(\gamma) \definedby \tsem{\V}(\gamma)
\\
   \tsem{\return\V}(\gamma) \definedby \return(\tsem\V(\gamma))
     &
\tsem{\Let\x\M\N}(\gamma) \definedby \tsem{\M}(\gamma) \bind \lambda
     a.\tsem{\N}(\gamma[\x \mapsto a])
\\
     \tsem{\abst\x\M}(\gamma) \definedby
     \lambda a.\tsem{\M}(\gamma[\x \mapsto a])
&
     \tsem{\apply\M\V}(\gamma) \definedby
     (\tsem\M(\gamma)) (\tsem\V(\gamma))
\\
     \tsem{\cpair{\M_1}{\M_2}}(\gamma) \definedby
     \pair{\tsem{\M_1}(\gamma)}
          {\tsem{\M_2}(\gamma)}
     &
     \tsem{\proj{i}{\M}}(\gamma) \definedby \projection_i(\tsem\M(\gamma))
\end{array}
\)
\end{tabular}
}

\newcommand\cbpvKindSystem{
  \def\SharedSpace{12pt}
\begin{tabular}{l@{\!\!}l}
  \begin{tabular}[t]{l}
  \figheading{Effect kinding} $\quad \boxed{\kinf\e{\EffectKind}}$
  \hspace{1cm}
  \(
  \overline{\kinf\emptyset\EffectKind}
  \)
  \bigskip\\
  \figheading{Value kinding} $\quad \boxed{\kinf\A\ValueKind}$\bigskip\\
  \multicolumn{1}{c}{\(
  \begin{array}{c}
    \inferrule{\tyv \in \Ty}{\kinf\tyv\ValueKind}
    \quad
    \inferrule{~}{\kinf\vUnitType\ValueKind}
    \quad
    \inferrule{\kinf{\A_1}\ValueKind\quad\kinf{\A_1}\ValueKind}{\kinf{\vProdType{\A_1}{\A_2}}{\ValueKind}}
    \\\\
    \inferrule{\text{for every $1 \leq i \leq n$: } \kinf{\A_i}\ValueKind}
              {\kinf{\VariantType{\VariantConstructor{\ell_1}{\A_1}
                    \vand\ldots\vand
                        \VariantConstructor{\ell_n}{\A_n}}}{\ValueKind}}
    \quad
    \inferrule{\kinf\C{\ComputationKind\e}}{\kinf{\U\e\C}\ValueKind}
  \end{array}
  \)
  }
\end{tabular}
&
  \begin{tabular}[t]{l}
  \figheading{Computation kinding}
    \begin{array}[t]{l}
    $\boxed{\kinf\C{\ComputationKind\e}}$ \\[1ex]
    (\kinf\e\EffectKind) \\
    \end{array} \bigskip\\
  \multicolumn{1}{c}{\(
  \begin{array}{c}
    \inferrule{
      \kinf\A\ValueKind\\
    }{\kinf{\F\A}{\ComputationKind\e}}
    \\[\SharedSpace]
    \inferrule{
      \kinf\A\ValueKind\\
      \kinf\C{\ComputationKind\e}
    }{\kinf{\Fun\A\C}{\ComputationKind\e}}
    \\[\SharedSpace]
    \inferrule{
      \kinf{\C_1}{\ComputationKind\e}\\
      \kinf{\C_2}{\ComputationKind\e}
    }{\kinf{\cProdType{\C_1}{\C_2}}{\ComputationKind\e}}
  \end{array}
  \)
  }
\end{tabular}
\end{tabular}
}

\newcommand\cbpvTypeSystem{
\begin{tabular}{l@{}l}
~\\[-1ex]
\begin{tabular}[t]{l}
  \figheading{Context kinding} $\quad \boxed{\kinf\G{\ContextKind}}$
    \\[\SharedSpace]
    \multicolumn{1}{c}{
    \(
    \inferrule{\text{for all $x \in \Dom\G$: } \kinf{\G(x)}\ValueKind}{\kinf \G\ContextKind}
  \)}
    \\[\SharedSpace]
  \figheading{Value typing} $\quad \boxed{\Ty\vinf\G\V\A}\quad $ \bigskip\\
  \multicolumn{1}{r}{($\kinf\G\ContextKind,\A\of\ValueKind$)}\\
  \multicolumn{1}{c}{
  \(
  \begin{array}{c}
    \inferrule
    {(x\of\A) \in \G}
    {\Ty\vinf\G\x\A}
    \qquad
    \inferrule
    {~}
    {\Ty\vinf\G{\vunit}{\vUnitType}}
    \\[\SharedSpace]
    \inferrule
    {\Ty\vinf\G{\V_1}{\A_1} \\ \Ty\vinf\G{\V_2}{\A_2}}
    {\Ty\vinf\G{\vpair{\V_1}{\V_2}}
            {\vProdType{\A_1}{\A_2}}}
    \\[\SharedSpace]
    \inferrule{\Ty\vinf\G\V{\A_i}}
    {\Ty\vinf\G{\inj{\ell_i}\V}{\begin{array}[t]{@{}l@{}}
                                  \VariantType{\VariantConstructor{\ell_1}{\A_1}\\
                                  \vand\ldots\\
                                  \vand
                        \VariantConstructor{\ell_n}{\A_n}}\end{array}}}
    \\[\SharedSpace]
    \inferrule
    {\Ty\cinf\G\M\C}
    {\Ty\vinf\G{\thunk\M}{\U{\E}\C}}

    \sharedValueTyping
  \end{array}
  \)
  }
\end{tabular}
&
\def\SharedSpace{12pt}
\begin{tabular}[t]{l}
  \figheading{Computation typing}
   \begin{array}[t]{l}
   \boxed{\Ty\cinf\G\M\C} \\[1ex]
   (\kinf\G\ContextKind,\E \of \EffectKind, \C\of\ComputationKind\E)\\
   \end{array} \bigskip\\
  \multicolumn{1}{r}{
  \(
  \begin{array}{c}
    \inferrule
    {\Ty\vinf\G{\V}{\vProdType{A_1}{A_2}} \quad
     \Ty\cinf{\G,\x_1\of\A_1,\x_2\of\A_2}{\M}\C}
    {\Ty\cinf\G{\Split{\V}{\x_1}{\x_2}{\M}}\C}
    \quad
    \inferrule
    {\Ty\vinf\G\V{\U{\E}\C}}
    {\Ty\cinf\G{\force\V}\C}
    \\[\SharedSpace]
    \inferrule
        {\Ty\vinf\G\V{}{\VariantType{\VariantConstructor{\ell_1}{\A_1}\vand\cdots\vand\VariantConstructor{\ell_n}{\A_n}}}
          \\\\
          \text{for every $1 \leq i \leq n$: } \Ty\cinf{\G, \x_i : \A_i}{\M_i}{\C}}
    {\Ty\cinf\G
      {
        \Variant{\V}{
          \VariantCase{\ell_1}{\x_1}{\M_1};
          \cdots;
          \VariantCase{\ell_n}{\x_n}{\M_n}
        }
      }\C}
    \\[\SharedSpace]
    \inferrule
    {\Ty\vinf\G\V\A}
    {\Ty\cinf\G{\return\V}{\F\A}}
    \qquad
    \inferrule
    {\Ty\cinf\G{\M}{\cProdType{\C_1}{\C_2}}}
    {\Ty\cinf\G{\proj{i}{\M}}\C_i}
    \\[\SharedSpace]
    \inferrule
    {\Ty\cinf\G\M{\F\A}  \\  \Ty\cinf{\G,\x\of\A}\N\C}
    {\Ty\cinf\G{\Let\x\M\N}\C}
    \quad
    \inferrule
    {\Ty\cinf{\G,\x\of\A}\M\C}
    {\Ty\cinf\G{\abst\x\M}{\Fun\A\C}}
    \\[\SharedSpace]
    \inferrule
    {\Ty\cinf\G\M{\Fun\A\C} \quad \Ty\vinf\G\V\A}
    {\Ty\cinf\G{\apply\M\V}\C}
    \\[\SharedSpace]
    \inferrule
    {\Ty\cinf\G{\M_1}{\C_1} \quad \Ty\cinf\G{\M_2}{\C_2}}
    {\Ty\cinf\G{\cpair{\M_1}{\M_2}}
            {\cProdType{\C_1}{\C_2}}}
  \end{array}
  \)
  }
  \end{tabular}
  \end{tabular}
}
 \newcommand\lameffSyntax{
  \hspace{-1em}
\begin{newsyntax}
\sharedSyntax{\ldots}{
       &\gor& {\operation\op\V}       & \synname{operation call}    \\
  &\gor& {\handle\M\H}} \        & \synname{handling construct}
\\
    &\mathrlap{\H\gdefinedby}& &\syncat{handlers}                             \\
       &    & \mathrlap{\set{\returnClause\x\M}}\hphantom{\mspace{120mu}a} & \synname{return clause}     \\
       &\gor& \mathrlap{\H \uplus \set{\handlerEntry{\op}{\p}{\k}{\N}}} & \synname{operation clause}
\end{newsyntax}
}

\newcommand\lameffTypes{
      \figheading{Kinds and types}\\
      \begin{newsyntax}
        {}&\mathrlap{\e  \gdefinedby  \ldots}&&   \syncat{effects}                               \\
        {}&\gor& \set{\op \Arity{\Ar}{\Pa}} \uplus \e & \synname{arity assignment}
        \\
        {}&\mathrlap{K\gdefinedby \ldots} && \syncat{kinds}                                   \\
        {}&\gor& \HandlerKind &\synname{handlers}                                 \\
        {}&\mathrlap{\HT  \gdefinedby {\HandlerType\A\e\C{\e'}}}&\hphantom{\mspace{120mu}a} & \syncat{handler types}
      \end{newsyntax}
}

\newcommand\lameffTypeSystemComputations{
        \figheading{Computation typing} $\quad\cdots$\\
        \multicolumn{1}{@{}c@{}}{
          \(\begin{array}{@{}c@{}}
              \inferrule
              {(\op\Arity\Ar\Pa) \in \e \\ \Ty\vinf\G\V\Pa}
              {\Ty\einf\e\G{\operation\op\V}{\F\Ar}}
    \\
    \inferrule
    {\Ty\einf\e\G\M{\F\A} \\
      \Ty\hinf\G\H{\HandlerType\A\e\C{\e'}}}
    {\Ty\einf{\e'}\G{\handle\M\H}\C }
  \end{array}
  \)}
}

\newcommand\lameffKindSystem{
  \figheading{Effect kinding} \(\quad
     \cdots\)\\[\SharedSpace]
     \multicolumn{1}{@{}c@{}}{
     \(
     \inferrule{
     \kinf\A\ValueKind\\
     \kinf\B\ValueKind\\
     \op\notin\e\\
     \kinf\e\EffectKind}
     {\kinf{\set{\op \Arity{\Ar}{\Pa}} \uplus \e}\EffectKind}
     \)}
     \medskip\\[\SharedSpace]
           \figheading{Handler kinding} $\quad\boxed{\kinf R\HandlerKind}$ \\[\SharedSpace]
     \multicolumn{1}{@{}c@{}}{
      \(
      \inferrule{
        \kinf\A\ValueKind\ \
        \kinf{\e, \e'}\EffectKind\ \
        \kinf\C{\ComputationKind{\e'}}
      }{
        \kinf{\HandlerType\A\e\C{\e'}}\HandlerKind
    }
     \)}
}

\newcommand\lameffHandlerTypingCaption{
  \figheading{Handler typing} $\ \boxed{\Ty\hinf\G\H R}\ $
                                          ($\kinf\G\ContextKind,R\of\HandlerKind$)
}

\newcommand\lameffHandlerTypingBody{
  \(
   \inferrule
   {
     \Ty\einf{\e}{\G,x\of\A}\M\C
     \\
     \text{for all ${1 \leq i \leq n}$: }\quad
     \Ty\einf{\e}{\G,\p\of\Pa_i,\k\of
       \U{\e}{(\Fun{\Ar_i}\C)}}{\N_i}\C
   }
   {
       \Ty\hinf\G
       {\set{\returnClause{\x}{\M}} \uplus {\set{\op_i\ \p\ \k \mapsto \N_i\suchthat 1 \leq i \leq n}}}
       {\HandlerType\A{\set{\op_i : \A_i \to \B_i \suchthat 1 \leq i \leq n}}\C{\e}}
   }
   \)
}

\newcommand\lameffKindingAndTypingExtension{
  \begin{tabular}{@{}l@{\quad}l@{}}
    \begin{tabular}[t]{@{}l@{}}
      \lameffTypes\medskip\\
      \lameffTypeSystemComputations\medskip\\
    \end{tabular}
 &
   \def\SharedSpace{15pt}
   \begin{tabular}[t]{@{}l@{}}
     \lameffKindSystem
   \end{tabular}\\
    \multicolumn{2}{@{}l@{}}\lameffHandlerTypingCaption\\[4pt]
    \multicolumn{2}{@{}c@{}}\lameffHandlerTypingBody
  \end{tabular}
}

\newcommand\lameffOpSemantics{
  \begin{tabular}[t]{l}
  \figheading{Frames and contexts}\\
  \begin{newsyntax}
      \quad\cdots\quad
     &\EF  \gdefinedby{} \ldots \gor \handle \hole \H& & \syncat{computation frame}
  \end{newsyntax}
    \\
  \figheading{Beta reduction}
    \\
  \begin{reductions}
     (\handleReturn) &
    \handle{&(\return\V)&}\H
    \breducesto
    \substitute{\handlerRet\H}{\for\V\x}
    \\
 (\handleOp) &
     \handle{&\CF[{\operation{\op}\V}]&}{\H}\breducesto\\
     &&\multicolumn{3}{r}{
       \substitute{\handlerOp\H\op}{
        \for{\V}{\p},
        \for{\thunk{\abst\x{\handle{\CF[\return \x]}{\H}}}}{\k}}
     }
  \end{reductions}
  \end{tabular}
}

\newcommand{\lameffTermDenSem}{
  \begin{tabular}[t]{@{}l@{}}
  \figheading{Computation terms} $\quad\cdots$\\
  \(
  \begin{array}[t]{@{}l}
    \tsem{\operation\op\V}(\gamma) \definedby \op_{\tsem{\V}\g}\seq[a
    \in \ksem \Ar]{\return a}
    \\
    \tsem {\handle\M\H}(\gamma) \definedby \tsem\M(\gamma) \bind f
    \\
    \multicolumn{1}{@{}r@{}}{\text{
        where $\sem \H(\gamma) = \pair{D}{f : \sem A \to \carrier{\sem C}}$
    }}
  \end{array}
  \)
  \end{tabular}
  &
  \begin{tabular}[t]{@{}l@{}}
  \figheading{Handler terms}\quad\\
  \(
    \tsem{\set{\returnClause{\x}{\M}} \uplus \seti[\op]{\op\ \p\ \k
        \mapsto \N_{\op}}} (\gamma) \definedby \pair{D}{f}
  \)\\ where $D$'s algebra structure and $f$ given by:
  \\
  \(\begin{array}[t]{@{}l@{\ \quad}r@{}}
    \sem{\op_q}_{D}\seq[a]{\xi_a} \definedby \tsem
    {\N_{\op}}(\substitute\gamma{\for{q}\p, \for{\seq[a]{\xi_a}}{k}}
    &
    f (a) \definedby \tsem\M(\substitute\gamma{\for ax})
  \end{array}
  \)
  \end{tabular}
}

\newcommand{\lameffDenSem}{
  \begin{tabular}[t]{@{}l@{}}
    \figheading{Effects} \\
    \(
    \ksem{\E} \definedby \T_{\set{\op_p : \ksem{\A} \suchthat (\op \Arity\Ar\Pa) \in \E, p \in \ksem\Pa}}
    \)
  \end{tabular}
  &
  \begin{tabular}[t]{@{}l@{}}
    \figheading{Handler types} \\
    \(\begin{array}[t]{@{}l@{}}
        \sem{{\HandlerType\A\e\C{\e'}}} \definedby
        \set{\text{$\sem\e$-algebras with carrier
        $\carrier{\sem\C}$}}\times \carrier{\sem{\C}}^{\sem\A}
      \end{array}
    \)
  \end{tabular}
}

\newcommand\lammonSyntax{
  \begin{newsyntax}
    ~&\mathrlap{\Monad \gdefinedby{}}&   & \syncat{monads} \\
                                     &\effdef*{\x&{}}\M {&\synname{return clause}\\
                                    &\mspace{65mu}\y} {f&{}}\N & \synname{bind clause}
\\
\sharedSyntax{\ldots}{
  &\gor& {\reflect{N}}                       &\synname{reflect}            \\
  &\gor& {\reify{N}}                         & \synname{reify}
}
\end{newsyntax}}

\newcommand{\lammonTypes}{
  \figheading{Kinds and types}\\
  \begin{newsyntax}
  {}&\mathrlap{\E    \gdefinedby{} \ldots} &&  \!\!\!\syncat{effects}\\
    &\gor &\effType \E\alpha C\Monad& \!\!\!\synname{layered monad}
  \end{newsyntax}
}

\newcommand\lammonMonadTypingCaption{\figheading{Monad typing} $\quad \boxed{\Ty\minf\Monad\E}\quad$}
\newcommand\lammonMonadTypingBody{
    \inferrule
        { \Ty,\alpha\einf\E{
            \x : \alpha
          }{\N_u}{\C}\quad
          \Ty,\alpha,\beta\einf\E{
            \y : {\U{\E}\C},
            f  : \U{\E}(\alpha \to \substitute\C{\for{\beta}{\alpha}})
          }{\N_b}{
            \substitute\C{\for{\beta}{\alpha}}
         }
        }
        {
          \begin{aligned}[t]
            \Ty&\minf* {\effdef*\x{N_u}\y f{N_b}}{\\&\quad\E}\alpha\C
          \end{aligned}
        }
}
\newcommand\lammonComputationTypingCaption{\figheading{Computation typing} $\quad\cdots\quad$}
\newcommand\lammonComputationTypingBody{
  \begin{mathpar}
    \inferrule
        {\Ty\einf\E\G\N{\substitute\C{\for{\A}{\alpha}}}}
        {\Ty\einf{\effType\E\alpha\C {\Monad}}\G{\reflect N}{\F A}}

    \inferrule
        {\Ty\minf\Monad{\effType\E\alpha\C \Monad}\\\\
         \Ty\einf{\effType\E\alpha\C \Monad}\G{\N}{\F A}
        }
        {\Ty\einf\E\G{\reify N}{\substitute{\C}{\for A\alpha}}{}}
   \end{mathpar}
}

\newcommand\lammonKindSystem{
  \figheading{Effect kinding}$\quad\cdots\ $\\
    \inferrule{
      \kinf[\Theta, \alpha ] C{\ComputationKind\E}\
      \minf{\Monad}
           {\effType \E\alpha C\Monad}
    }{
      \kinf{\effType \E\alpha C\Monad}\EffectKind}
}

\newcommand\lammonKindingAndTypingExtension{
  \begin{tabular}{@{}l@{\quad}l@{}}
    \begin{tabular}[t]{@{}l@{}}
      \lammonTypes\\
    \end{tabular}
 &
   \def\SharedSpace{15pt}
   \begin{tabular}[t]{@{}l@{}}
     \lammonKindSystem
   \end{tabular}\\
    \multicolumn{2}{l}~\\[0pt]
    \multicolumn{2}{@{}l@{}}{\lammonMonadTypingCaption
    \lammonMonadTypingBody}   \\[4pt]
    \multicolumn{2}{@{}l@{}}\lammonComputationTypingCaption\\[4pt]
    \multicolumn{2}{@{}c@{}}\lammonComputationTypingBody   \\[4pt]
  \end{tabular}
}

\newcommand{\lammonOpSemantics}{
\begin{tabular}[t]{l}
\figheading{Frames and contexts}\\
\(
    \quad\cdots\quad  \EF   \gdefinedby{} \EB \gor \reify{[~]}  \syncat{computation frames}     \quad\cdots
\)
\\
\figheading{Beta reduction}$\quad\cdots$ for every $\Monad = \effdef{\abst\x\N_u}{\abst\y{\abst f{\N_b}}}$:\\
\begin{reductions}
  (\reifyRule) &
  \reify{&\return \V&}
  \breducesto \substitute{\N_u}{\for \V\x}
  \\
  (\reflectRule)
  &
  \reify{& \CF[\reflect{\N }]&}
  \breducesto\\
  &&\mathrlap{\substitute{\N_b}{\for{\thunk{\N}}\y, \for{\thunk{(\lambda{x.}{\reify{\CF[\return x]}})}}f}}
\end{reductions}
\end{tabular}
}

\newcommand{\lammonDenSem}{
\figheading{Effects} $\quad\cdots$\\
\(
\ksem{\effType\E\alpha\C {N_u}{N_b}}
\definedby \triple{T}{\return{}}{\bind}
\)
\\where
\(
\begin{array}[t]{l@{}}
T X \definedby \carrier{\tsem[{(\theta[\alpha \mapsto X])}]{C}}
\quad
\return{}^X \definedby \tsem[{(\theta[\alpha\mapsto X])}]{N_u} : X \to T  X
\\
  \bind^{X,Y} \definedby \tsem[{(\theta[\alpha_1 \mapsto X, \alpha_2 \mapsto Y])}]{N_b} : T X \to (X \to T Y) \to T Y \\
\text{(provided these form a monad)} \\
\end{array}\)\\
}

\newcommand{\lammonDenSemTerms}{
  \figheading{Monads}\\
  \(
  \sem{\Ty\minf \Monad\E} \definedby \sem{\E}
  \)
  \\[4pt]
  \figheading{Computation terms} $\quad\cdots$\\
  \(\begin{array}[t]{@{}l@{}l@{}}
    \sem{\reify{N}}&(\gamma) \definedby \sem{N}(\gamma)
    \\
    \sem{\reflect{N}\ }&(\gamma) \definedby \sem{N}(\gamma)
    \end{array}
    \)
}

 \newcommand\lamdelSyntax{
  \begin{newsyntax}
    \sharedSyntax{\ldots}{
    &\gor& {\ShiftZ kM} & \synname{shift-$0$} \\
    &\gor& {\Reset MxN} & \synname{reset}
  }
  \end{newsyntax}
}

\newcommand\lamdelKindingAndTypingExtension{
  \begin{tabular}{@{}l@{\qquad\qquad}l@{}}
    \begin{tabular}[t]{@{}l@{}}
      \lamdelTypes\\
    \end{tabular}
 &
   \def\SharedSpace{15pt}
   \begin{tabular}[t]{@{}l@{}}
     \lamdelKindSystem
   \end{tabular}\\
    \multicolumn{2}{l}~\\[0pt]
    \multicolumn{2}{@{}l@{}}\lamdelComputationTypingCaption\\[4pt]
    \multicolumn{2}{@{}c@{}}\lamdelTypeSystem              \\[4pt]
  \end{tabular}
}

\newcommand\lamdelOpSemantics{
  \begin{tabular}[t]{l}
  \figheading{Frames and contexts}\\
  \begin{newsyntax}
      \quad\cdots\quad
     \EF & \gdefinedby \ldots & \gor \Reset \hole xN & \syncat{computation frame}
  \end{newsyntax}
  \\
  \figheading{Beta reduction}\\
     $\quad\cdots\quad$
  \begin{reductions}
 (\resetRule) &
    \Reset{(\return\V)}xM
      &\breducesto& \substitute\M{\for\V\x}
  \\
 (\shiftRule) &
      \Reset{\CF[{\ShiftZ kM}]}xN
       &\breducesto&
      \substitute{\M}{
        \for {\abst y{\Reset{\CF[\return y]}xN}}
             {k}}
 \end{reductions}
\end{tabular}
}

\newcommand\lamdelTypes{
  \figheading{Kinds and types}\\
  \begin{newsyntax}
    {} & \mathrlap{\E\gdefinedby{} \ldots} &\hphantom{control}& \syncat{effects} \\
       &\gor &\E,\C & \synname{enclosing continuation type}
  \end{newsyntax}
}

\newcommand\lamdelKindSystem{
  \figheading{Effect kinding} $\quad \cdots\quad$\\
  \inferrule{
    \kinf\E\EffectKind
    \\
    \kinf\C{\ComputationKind\E}
    }
    {\kinf{\E,C}\EffectKind}

}

\newcommand\lamdelComputationTypingCaption{  \figheading{Computation typing} $\quad\ldots$}

\newcommand\lamdelTypeSystem{
  \begin{mathpar}
     \inferrule
     {\Ty\einf\E{\G,k:\U{\E}(A\to \C)} M{\C}}
     {\Ty\einf{\E,C}\G{\ShiftZ kM}{\F A}}
    \quad
    \inferrule
     {\Ty\einf{\E,\C}\G M{\F A}\\
      \Ty\einf\E{\G,x:A}NC
     }
     {\Ty\einf{\E}\G{\Reset MxN}C}
  \end{mathpar}
}

\newcommand\E{E}
\newcommand\Monad{T}
\newcommand\pure{\emptyset}
\newcommand\tyv{\alpha}

\newcommand\Ty{\Theta}
\newcommand\tysem{\theta}

\newcommand\bit{\mathbf{bit}}

\DeclareMathDelimiter{\ullcorner}{\mathopen}{okMnSymbolLargeSymbols}{'62}{okMnSymbolLargeSymbols}{'62}
\DeclareMathDelimiter{\ulrcorner}{\mathclose}{okMnSymbolLargeSymbols}{'67}{okMnSymbolLargeSymbols}{'67}

\newcommand\trans[2][]{\underline{#2}}

\newcommand\macro[2]{#1\mathord{\rightarrow}#2}
\renewcommand\explain[2]{&\{\text{#2}\}\\&#1}
\newcommand\Cont{\op[Cont]}
\newcommand\Reader{\op[Reader]}

\newcommand\diverges{\reducesto^\infty}
\usepackage{centernot}
\newcommand\ndiverges{{\centernot\reducesto}^\infty}
\newcommand\PluggedContexts[1]{\Xi[{#1}]}
\newcommand\eqcong{\simeq_{\mathrm{cong}}}

\newcommand\mb{\varstyle{b}}
\newcommand\mx{\varstyle{x}}
\newcommand\tru{\texttt{True}}
\newcommand\fls{\texttt{False}}
\WithSuffix\newcommand\Let*[2]{#1;\ #2}

\usepackage{booktabs}

\acmJournal{PACMPL}
\acmVolume{1}
\acmNumber{1}
\acmArticle{39}
\acmYear{2017}
\acmMonth{3}
\acmArticleSeq{1}

\setcopyright{rightsretained}

\acmDOI{0000001.0000001}

\received{February 2017}
 \author{Yannick Forster}
\email{forster@ps.uni-saarland.de}
\affiliation{
  \institution{Saarland University}
  \streetaddress{Saarland Informatics Campus}
  \city{Saarbr\"ucken}
  \postcode{66123}
  \country{Germany}}
\affiliation{
  \institution{University of Cambridge}
  \department{Computer Laboratory}
  \streetaddress{William Gates Building, 15 JJ Thomson Avenue}
  \city{Cambridge}
  \postcode{CB3 0FD}
  \country{England}}
\author{Ohad Kammar}
\orcid{0000-0002-2071-0929}
\email{ohad.kammar@cs.ox.ac.uk}
\affiliation{
  \institution{University of Oxford}
  \department{Department of Computer Science}
  \streetaddress{Wolfson Building, Parks Road}
  \city{Oxford}
  \postcode{OX1 3QD}
  \country{England}}
\affiliation{
  \institution{University of Cambridge}
  \department{Computer Laboratory}
  \streetaddress{William Gates Building, 15 JJ Thomson Avenue}
  \city{Cambridge}
  \postcode{CB3 0FD}
  \country{England}}
\author{Sam Lindley}
\email{sam.lindley@ed.ac.uk}
\affiliation{
  \institution{University of Edinburgh}
  \department{School of Informatics}
  \streetaddress{Informatics Forum, 10 Crichton Street}
  \city{Edinburgh}
  \postcode{EH8 9AB}
  \country{Scotland}}
\author{Matija Pretnar}
\email{matija.pretnar@fmf.uni-lj.si}
\affiliation{
  \institution{University of Ljubljana}
  \department{Faculty of Mathematics and Physics}
  \streetaddress{Jadranska 21}
  \city{Ljubljana}
  \postcode{SI-1000 }
  \country{Slovenia}
}

\makeatletter
\begin{document}
\title[On the Expressive Power of User-Defined Effects]
      {On the Expressive Power of User-Defined Effects:}
\subtitle{Effect Handlers, Monadic Reflection, Delimited Control}

\begin{abstract}
  We compare the expressive power of three programming abstractions
  for user-defined computational effects: Bauer and Pretnar's effect
  handlers, Filinski's monadic reflection, and delimited control
  without answer-type-modification. This comparison allows a precise
  discussion about the relative expressiveness of each programming
  abstraction. It also demonstrates the sensitivity of the relative
  expressiveness of user-defined effects to seemingly orthogonal
  language features.

  We present three calculi, one per abstraction, extending Levy's
  call-by-push-value. For each calculus, we present syntax,
  operational semantics, a natural type-and-effect system, and, for
  effect handlers and monadic reflection, a set-theoretic denotational
  semantics. We establish their basic meta-theoretic properties:
  safety, termination, and, where applicable, soundness and
  adequacy. Using Felleisen's notion of a macro translation, we show
  that these abstractions can macro-express each other, and show which
  translations preserve typeability.  We use the adequate finitary
  set-theoretic denotational semantics for the monadic calculus to
  show that effect handlers cannot be macro-expressed while preserving
  typeability either by monadic reflection or by delimited control. We
  supplement our development with a mechanised Abella formalisation.
\end{abstract}


 
\maketitle

\section{Introduction}

How should we compare abstractions for user-defined effects?

The use of computational effects, such as file, terminal, and network
I/O, random-number generation, and memory allocation and mutation, is
controversial in functional programming. While languages like Scheme
and ML allow these effects to occur everywhere, pure languages like
Haskell restrict the use of effects. A main trade-off when
incorporating computational effects into the language is giving up
some of the most basic properties of the lambda calculus, like
$\beta$-equality, referential transparency, and confluence. The loss
of these properties may lead to unpredictable behaviour in lazy
languages like Haskell, or limit the applicability of correctness
preserving transformations like common subexpression elimination or
code motion.

\emph{Monads}~\citep{Moggi89, spivey:a-functional-theory-of-exceptions, wadler:comprehending-monads} are the established
abstraction for incorporating effects into pure languages. The introduction
of monads into Haskell led to their additional use as a programming
abstraction, allowing new effects to be declared and used as if they
were native. Examples include
parsing~\citep{hutton-meijer:monadic-parsing-in-haskell}, backtracking
and constraint solving~\citep{schrijvers-et-al:search-combinators},
and mechanised reasoning~\citep{ziliani-et-al:mtac,
  bulwahn-et-al:imperative-functional-programming-with-isabelle/hol}.
Libraries now exist for monadic programming even in impure languages
such as
OCaml\footnote{\url{http://www.cas.mcmaster.ca/~carette/pa_monad/}},
Scheme\footnote{\url{http://okmij.org/ftp/Scheme/monad-in-Scheme.html}},
and
C++~\citep{sinkovics-zoltan:implementing-monads-for-c++-template-metaprograms}.

\citet{bauer-pretnar:programming-with-algebraic-effects-and-handlers}
propose the use of \emph{algebraic effects and handlers} to structure
programs with user-defined effects. In this approach, the programmer
first declares \emph{algebraic operations} as the syntactic constructs
she will use to cause the effects, in analogy with declaring new
exceptions. Then, she defines \emph{effect handlers} that describe how
to handle these operations, in analogy with exception handlers. While
exceptions immediately transfer control to the enclosing handler
without resumption, a computation may continue in the same position
following an effect operation.
In order to support resumption, an effect handler has access to the
\emph{continuation} at the point of effect invocation. Thus algebraic
effects and handlers provide a form of \emph{delimited control}.

Delimited control operators have long been used to encode
effects~\citep{danvy-advanced-thesis} and algorithms with
sophisticated control
flow~\citep{felleisn-wand-friedman-duba:abstract-continuations}. There
are many variants of such control operators, and their
inter-relationships are
subtle~\citep{shan:a-static-simulation-of-dynamic-delimited-control},
and often appear only in folklore. Here we focus on a specific such
operator: \emph{shift-zero and dollar without
  answer-type-modification}~\citep{materzok-biernacki:a-dynamic-interpretation-of-the-CPS-hierarchy},
whose operational semantics and type system are the closest to effect
handlers and monads.

We study these three different abstractions for user-defined effects:
effect handlers, monads, and delimited control operators. Our goal is
to enable language designers to conduct a precise and informed
discussion about the relative expressiveness of each abstraction. In
order to compare them, we build on an idealised calculus for
functional-imperative programming, namely
call-by-push-value~\citep{levy:book}, and extend it with each of the
three abstractions and their corresponding natural type systems. We
then assess the expressive power of each abstraction by rigorously
comparing and analysing these calculi.

We use Felleisen's notion of macro expressibility~\citep{\felleisen}:
when a programming language $\lang$ is extended by some feature, we
say that the extended language $\lang_+$ is \emph{macro expressible}
when there is a syntax-directed translation from $\lang_+$ to $\lang$
that keeps the features in $\lang$ fixed. Felleisen introduces this
notion of reduction to study the expressive power of Turing-complete
calculi, as macro expressivity is more sensitive in these contexts
than computability and complexity notions of reduction.  We adapt
Felleisen's notion to the situation where one extension $\lang_+^1$ of
a base calculus $\lang$ is macro expressible in another extension
$\lang_+^2$ of the same base calculus $\lang$. Doing so enable us to
formally compare the expressive power for each approach to
user-defined effects.

In the first instance, we show that, disregarding types, all three
abstractions are macro-expressible in terms of one another, giving six
macro-expression translations. Some of these translations are known in
less rigorous forms, either published, or in folklore. One
translation, macro-expressing effect-handlers in delimited control,
improves on previous concrete
implementations~\citep{kammar-lindley-oury:handlers-in-action}, which
rely on the existence of a global higher-order memory cell storing a
stack of effect-handlers. The translation from monadic reflection to
effect handlers is completely novel.

We also establish whether these translations preserve typeability: the
translations of some well-typed programs are untypeable. We show that
the translation from delimited control to monadic reflection preserves
typeability.
A potential difference between the expressive power of handler type
systems and between monadic reflection and delimited control type
systems was recently pointed out by
\citet{kammar-pretnar:no-value-restriction-is-needed-for-algebraic-effects-and-handlers},
who give a straightforward typeability preserving macro-translation of
\emph{delimited dynamic state} into a calculus effect handlers,
whereas existing translations using monads and delimited control
require more
sophistication~\cite{kiselyov-shan-sabry:delimited-dynamic-binding}.
Here, we establish this difference: we demonstrate how to use the
denotational semantics for the monadic calculus to prove that there
exists no no macro translation from the effect handlers calculus to
the monadic reflection calculus that preserves typeability. This
set-theoretic denotational semantics and its adequacy for
\citepos*{filinski:monads-in-action}{multi-monadic metalanguage} is
another piece of folklore which we prove here. We conjecture that a
similar proof, though with more mathematical sophistication, can be
used to prove the non-existence of a typeability-preserving
macro-expression translation from the monadic calculus to effect
handlers.  To this end, we give adequate set-theoretic semantics to
the effect handler calculus with its type-and-effect system, and
highlight the critical semantic invariant a monadic calculus will
invalidate.

\figref{summary} summarises our contributions and conjectured
results. Untyped calculi appear on the left and their typed
equivalents on the right. Unlabelled arrows between the typed calculi
signify that the corresponding macro translation between the untyped
calculi preserves typeability. Arrows labelled by $*$ are new untyped
translations. Arrows labelled by $\nexists$ signify that \emph{no}
macro translation exists between the calculi, not even a partial macro
translation that is only defined for well-typed programs.

\begin{figure}
  \begin{center}
    \includegraphics{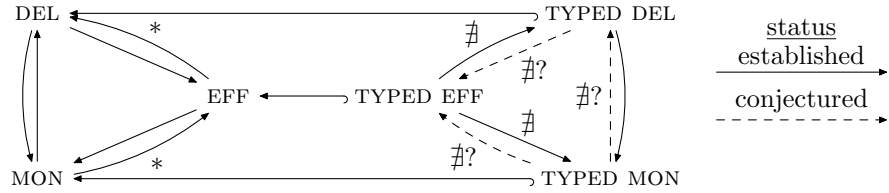}
  \end{center}
  \caption{Existing and conjectured macro translations}\figlabel{summary}
\end{figure}

The non-expressivity results are sensitive to the precise collection
of features in each calculus. For example, extending the base calculus
with inductive types and primitive recursion would create gaps in our
non-existence arguments, and we conjecture that extending the calculi
with various forms of polymorphism would make our untyped translations
typeability-preserving. Adding more features to each calculus blurs
the distinction between each abstraction. This sensitivity means that
in a realistic programming language, such as Haskell, OCaml, or
Scheme, the different abstractions are often practically
equivalent~\cite{schrijvers-pirog-wu-jaskelioff}. It also teaches us
that meaningful relative expressivity results \emph{must}
be stated within a rigorous framework such as a formal calculus, where
the exact assumptions and features are made explicit. The full picture
is still far from complete, and our work lays the foundation for such
a precise treatment.

We supplement our pencil-and-paper proofs with a mechanised
formalisation in the Abella proof
assistant~\citep{abella:system-description,gacek:thesis} of the more
syntactic aspects of our work. Specifically, we formalise a
\citepos-{wright-felleisen:a-syntactic-approach-to-type-soundness}{
  style progress-and-preservation soundness theorem}, which we also call
\emph{safety}, for each calculus, and correctness theorems for our
translations.

We make the following contributions:
\begin{itemize}
\item three formal calculi, i.e., syntax and semantics, for effect
  handlers, monadic reflection, and delimited control extending a
  shared call-by-push-value core, and their meta-theory:
  \begin{itemize}
  \item set-theoretic denotational semantics for effect
    handlers and monadic reflection;
  \item denotational soundness and adequacy proofs for effect handlers and monadic reflection;
  \item a termination proof for monadic reflection (proofs for the other calculi appear in existing work);
  \end{itemize}
\item six macro-translations between the three untyped calculi, and variations on three of those translations;

\item formally mechanised meta-theory in Abella\footnote{
  \url{https://github.com/matijapretnar/user-defined-effects-formalization}
}
  comprising:
  \begin{itemize}
  \item progress and preservation theorems;
  \item the translations between the untyped calculi; and
  \item their correctness proofs in terms of formal simulation results;
  \end{itemize}
\item typeability preservation of the macro translation from delimited
  control to monadic reflection;
  and
\item a proof that there exists no typeability-preserving macro
  translation from effect handlers to either monadic reflection or
  delimited control.
\end{itemize}

We structure the remainder of the paper as
follows. Sections~\ref{sec:cbpv}--
\ref{sec:del} present the core calculus and its extensions with
effect handlers, monadic reflection, and delimited control,
in this order, along with their meta-theoretic
properties. Section~\ref{sec:translations} presents the macro
translations between these calculi, their correctness, and
typeability-preservation. Section~\ref{sec:conclusion} concludes and
outlines further work.

 \section{The core-calculus: $\cbpv$}\label{sec:cbpv}
We are interested in a functional-imperative calculus where effects
and higher-order features interact well. Levy's call-by-push-value
(CBPV) calculus~\citep{levy:book} fits the bill. The CBPV
paradigm subsumes call-by-name and call-by-value, both syntactically
and semantically.  In CBPV evaluation order is explicit, and the way
it combines computational effects with higher-order features yields
simpler program logic reasoning
principles~\citep{plotkin-pretnar:a-logic-for-algebraic-effects,
  kammar-plotkin:algebraic-foundations-for-effect-dependent-optimisations}.
CBPV allows us to uniformly deal with call-by-value and call-by-name
evaluation strategies, making the theoretical development relevant to
both ML-like and Haskell-like languages.  We extend it with a
type-and-effect system, and, as \emph{adjunctions} form the semantic
basis for CBPV, we call the resulting calculus the
\emph{multi-adjunctive metalanguage} ($\cbpv$).

\begin{figure}
  \cbpvSyntax{}
  \caption{$\cbpv{}$ syntax}
  \figlabel{cbpv-syntax}
\end{figure}

\figref{cbpv-syntax} presents $\cbpv$'s raw term syntax, which
distinguishes between values (data) and computations (programs). We
assume a countable set of variables ranged over by $x$, $y$, $\ldots$,
and a countable set of variant constructor literals ranged over by
$\ell$. The unit value, product of values, and finite variants/sums
are standard.  A computation can be suspended as a thunk $\thunk M$,
which may be passed around.  Products and variants are eliminated with
standard pattern matching constructs. Thunks can be forced to resume
their execution. A computation may simply return a value, and two
computations can be sequenced, as in Haskell's {\tt do} notation. A
function computation abstracts over values to which it may be
applied. In order to pass a function $\abst x M$ as data, it must
first be suspended as a thunk $\thunk{\abst x M}$. For completeness,
we also include CBPV's binary computation products, which subsume
projections on product values in call-by-name languages.

\example Using the boolean values $\VariantConstructor\tru\vunit$
and $\VariantConstructor\fls\vunit$, we can implement a logical \emph{not}
operation:
\[
not = \thunk{
      \abst \mb
      \Variant {\begin{aligned}[t]\mb}{
       &\VariantCase{\mathrlap{\tru}\hphantom{\fls}}{\mx}{\return{\inj{\fls}\vunit}}
        \\
        &\VariantCase{\fls}{\mx}{\return{\inj{\mathrlap{\tru}\hphantom{\fls}}\vunit}}
      }
    }
        \end{aligned}
\]

\figref{cbpv-op-semantics} presents $\cbpv$'s standard structural
operational semantics, in the style of
\citet{felleisen-friedman:a-reduction-semantics-for-imperative-higher-order-languages}. In
order to reuse the core definitions as much as possible, we refactor
the semantics into $\beta$-reduction rules and a single congruence
rule. As usual, a $\beta$-reduction reduces a matching pair of
introduction and elimination forms. We specify in the definition of
evaluation contexts the \emph{basic frames}, which all our extensions
will share. We use $\hole$ to denote the hole in each frame or
context, which signifies which term should evaluate first, and define
substitution frames and terms for holes ($\EC[\EF\hole]$, $\EC[M]$) in
the standard way. Later, in each calculus we will make use of
\emph{hoisting frames} in order to capture continuations, stacks of
basic frames, extending from a control operator to the nearest
delimiter. As usual, a reducible term can be decomposed into at most
one pair of evaluation context and $\beta$-reducible term,
making the semantics deterministic.

\begin{figure}
  \cbpvOpSemantics{}
  \caption{$\cbpv{}$ operational semantics}
  \figlabel{cbpv-op-semantics}
\end{figure}

\example With this semantics we have
$\apply{\force{not}}
       {(\VariantConstructor\tru\vunit)} \reducesto+ \return {\VariantConstructor{(\fls\vunit)}}$.

In this development, we use the following standard syntactic sugar. We
use nested patterns in our pattern matching constructs. We abbreviate
the variant constructors to their labels, and omit the unit value,
e.g., $\tru$ desugars to $\VariantConstructor\tru\vunit$. We allow the
application of functions and the elimination constructs to apply to
arbitrary computations, and not just values, by setting for example
\( M\ N \definedby \Let\x\N{M\ x} \)
for some fresh $x$, giving a more readable, albeit call-by-value,
appearance.

\example As a running example we express boolean state in each of our calculi
such that we can write code like $\var{toggle}$ in \subfigref{state
  introduction}{idealised toggle} which toggles the state and returns
the value of the original state. In \cbpv, we do so via a standard
state-passing transformation, as in \subfigref{state
  introduction}{state passing}, and run $\var{toggle}$ with the
initial value $\tru$ to get the expected result \(
\apply{\apply{\force{\var{runState}}}{\var{toggle}}}{\tru}
\reducesto^\star \vpair{\tru}{\fls} \).
This transformation is \emph{not} a macro translation. In addition to
the definition of $\var{put}$ and $\var{get}$, it globally threads
the state through $\var{toggle}$'s structure.
In later section, each abstraction provides a different means for
macro-expressing state.

\begin{figure}[b]
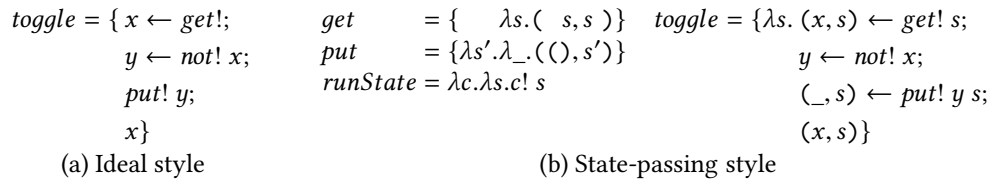

  \begin{tabular}{lr}
    \begin{tabular}[t]{c}
    \(
    \var{toggle} = \thunk{
      \begin{aligned}[t]
      &\Let{x}{\force{\var{get}}}\\
      &\Let{y}{\apply{\force{\var{not}}}{x}}\\
      &\Let*{\apply{\force{\var{put}}}
          y}\\
      & x
    }
      \end{aligned}
      \)
      \\
      \subcaptionn{Ideal style}{idealised toggle}
    \end{tabular}
    &
    \begin{tabular}[t]{c}
    \begin{tabular}[t]{cc}
      \(
    \begin{array}[t]{@{}l@{~}c@{~}l@{}l@{}r@{}l@{}r@{}}
      get &=& \thunk{&\abst{s}{\vpair {&s&}{s&}}} \\
      put &=& \thunk{\abst{s'}{&\abst{\_}{\vpair{&\vunit&} {s'&}}}} \\
      \var{runState} &=& \multicolumn{3}{@{}l@{}}{
        \abst c\abst s\apply {\force c}s
      }
    \end{array}
    \)
    &
    \(
    \var{toggle} = \thunk{\abst s
      \begin{aligned}[t]
      &\Let{\vpair xs}{\apply{\force{\var{get}}}s}\\
      &\Let{y}{\apply{\force{\var{not}}}{x}}\\
      &\Let{\vpair \_ s}{\apply{\apply{\force{\var{put}}}
          y}s}\\
      &{\vpair xs}
    }
    \end{aligned}
      \)
    \end{tabular}
      \\
      \subcaptionn{State-passing style}{state passing}
    \end{tabular}
  \end{tabular}
  \caption{User-defined boolean state}\figlabel{state introduction}
\end{figure}

\figref{cbpv-types} presents $\cbpv{}$'s types and effects. $\cbpv$ is
a variant of
\citepos*{kammar-plotkin:algebraic-foundations-for-effect-dependent-optimisations}{multi-adjunctive
  intermediate language} without effect operations or coercions. As a
core calculus for three calculi with very different notions of effect,
$\cbpv$ is pure, and the only shared effect is the empty effect
$\pure$. We include a kind system, unneeded in traditional CBPV where
a context-free distinction between values and computations forces
types to be well-formed. The two points of difference from CBPV are
the kind of effects, and the refinement of the computation kind by
well-kinded effects $\E$. The other available kinds are the standard
value kind and a kind for well-formed environments (without type
dependencies).  Our type system includes value-type variables (which
we will later use for defining monads parametrically).
Simple value types are standard CBPV value types, and each type of
thunks includes an effect annotation describing the effects of these
thunks. Computation types include returners $\F \A$, which are
computations that return a value of type $\A$, similar to the monadic
type $\keyword{Monad\ } m \implies m\ a$ in Haskell. Functions are
computations and only take values as arguments. We include CBPV's
computation products, which account for product elimination via
projection in call-by-name languages. To ensure the well-kindedness of
types, which may contain type-variables, we use type environments in a
list notation that denotes sets of type-variables. Similarly, we use a
list notation for value environments, which are functions from a
finite set of variable names to the set of \emph{value} types.

\begin{figure}
  \cbpvTypes{}
  \caption{$\cbpv{}$ kinds and types}
  \figlabel{cbpv-types}
\end{figure}

\example The type of booleans \( \bit \) is given by
\(\VariantType{\VariantConstructor{\op[False]}\vUnitType\vand\VariantConstructor{\op[True]}\vUnitType}
\).

\figref{cbpv-kind-and-type-system} presents the kind and type
systems. The only effect ($\pure$) is well-kinded. Type variables must
appear in the current type environment, and they are always value
types. The remaining value and computation types and environments have
straightforward structural kinding conditions.  Thunks of
$\E$-computations of type $\C$ require the type $\C$ to be
well-kinded, which includes the side-condition that $\E$ is a
well-kinded effect. This kind system has the property that each valid
kinding judgement has a unique derivation.
Value type judgements assert that a value term has a well-formed value
type under a well-formed environment in some type variable
environment. The rules for simple types are straightforward.
Observe how the effect annotation moves between the
$\E$-computation type judgement and the type of $\E$-thunks. The side
condition for computation type judgements asserts that a computation
term has a well-formed $\E$-computation type under a well-formed
environment for some well-formed effect $\E$ under some type variable
environment. The rules for variables, value and computation products,
variants, and functions are straightforward. The rules for thunking
and forcing ensure the computation's effect annotation agrees with the
effect annotation of the thunk.  The rule for $\return{}{}$ allows us
to return a value at any effect annotation, reflecting the fact that
this is a \emph{may}-effect system: the effect annotations track which
effects may be caused, without prescribing that any effect \emph{must} occur.  The
rule for sequencing reflects our choice to omit any form of effect
coercion, subeffecting, or effect polymorphism: the three effect
annotations must agree. There are more sophisticated effect systems
which allow more
flexibility~\citep{katsumata:parametric-effect-monads-and-semantics-of-effect-systems}. We
leave the precise treatment of such extensions to later work.

\begin{figure}
  \cbpvKindSystem{}
  \cbpvTypeSystem{}
  \caption{$\cbpv{}$ kind and type system}
  \figlabel{cbpv-kind-and-type-system}
\end{figure}

\example The values from \subfigref{state
  introduction}{state passing} have the following types:
\begin{mathpar}
  \var{not} : \U\pure(\Fun\bit{\F\bit})

  \var{get} : \U\pure(\Fun\bit \F{(\vProdType\bit \bit)})

  \var{put} : \U\pure(\Fun\bit {\Fun\bit \F{(\vProdType\bit \bit)}})

  \var{toggle} : \U\pure(\Fun\bit{\F(\vProdType\bit\bit)})

  \var{runState} : \U\pure(\Fun{\U\pure(\Fun\bit{\F(\vProdType\bit\bit)})}{\Fun\bit\F(\vProdType\bit\bit)})
\end{mathpar}

\begin{theorem}[\cbpv safety]\theoremlabel{cbpv safety}
  Well-typed programs don't go wrong: for all closed \cbpv returners
  $\Ty\einf\emptyset{}M{\F\A}$, either $M \reducesto N$ for some
  $\Ty\einf\emptyset{}{N}{\F\A}$ or else ${M = \return V}$ for some
  $\Ty\vinf {}V\A$.
\end{theorem}

The proof is standard and formalised in Abella, established by
inductive proofs of progress and preservation.

We extend existing termination results for
CBPV~\citep{Doczkal2007,Doczkal2009}. We say that a term $M$
\emph{diverges}, and write $M \diverges$ if for every
$n \in \naturals$ there exists some $N$ such that $M \reducesto^n
N$. We say that $M$ \emph{does not diverge} when
$M \ndiverges$.
\begin{theorem}[\cbpv termination]\theoremlabel{cbpv termination}
  There are no infinite reduction sequences: for all \cbpv terms
  $\cinf[\pure]{}M{\F\A}$, we have $M \not\diverges$, and there exists
  some unique $\vinf {}\V\A$ such that
  \( M \reducesto^\star \return \V \).
\end{theorem}
The proof uses
\citepos*{tait-intensional-interpretations-of-functionals-of-finite-type-i}{method}
to establish totality, defining a relational interpretation to types
and establishing a basic lemma, and the notion of lifting from
\citepos*{hermida:thesis}{thesis} to define the monadic lifting of a
predicate. The remainder of the proof is immediate as the semantics is
deterministic.

For the purpose of defining contextual equivalence, we define the
subclass of \emph{ground types}:
\begin{syntax}
  \groundTypes
\end{syntax}

The definition of program contexts $\Ctx\hole$ and their type
judgements is straightforward but tedious and lengthy with four kinds
of judgements, and so we take a different approach. Informally, given
two computation terms $M_1$ and $M_2$, in order to define their
contextual equivalence, we need to quantify over the set \(
{\PluggedContexts{M_1,M_2} \definedby \set{\pair{\Ctx[M_1]}{\Ctx[M_2]} \suchthat
  \text{$\Ctx\hole$ is a well-typed context}}} \). Once we define this
set, we do not need contexts, their type system, or their semantics in
the remainder of the development, and so we will define this set
directly.

We say that an environment $\G'$ \emph{extends} an environment $\G$,
and write $\G' \geq \G$ if $\G'$ extends $\G$ as a partial function
from identifiers to value types. Given two well-typed computations
${\geninf{\Ty_0}{\G_0}{\E_0} {M_1}{\C_0}}$ and
${\geninf{\Ty_0}{\G_0}{\E_0} {M_2}{\C_0}}$, let
$\PluggedContexts{\geninf{\Ty_0}{\G_0}{\E_0}{M_1,M_2}{\C_0}}$ be the
smallest set of tuples $\seq{\Ty', \G', V_1, V_2, \A}$ and
$\seq{\Ty', \G', \E', N_1, N_2, \C}$ that is compatible with the
typing rules and contains all the tuples
$\seq{\Ty, \G, \E_0, M_1, M_2, \C_0}$, where $\Ty \supset \Ty_0$ and
$\G \geq \G_0$. The tuples $\seq{\Ty', \G', V_1, V_2, \A}$ and
$\seq{\Ty', \G', \E', N_1, N_2, \C}$ represent
$\geninf{\Ty'}{\G'}{}{V_1,V_2}{\A}$ and
$\geninf{\Ty'}{\G'}{\E'}{N_1,N_2}{\C}$, respectively. The
compatibility with the rules means, for example, that if
$\seq{\Ty', \G', V_1, V_2, \A}$ is in
$\PluggedContexts{\geninf{\Ty_0}{\G_0}{\E_0}{M_1,M_2}{\C_0}}$, then so
is $\seq{\Ty', \G', \pure, \return {V_1}, \return {V_2}, \F\A}$.

If we do define program contexts $\Ctx\hole$, we
can then show that this set consists of all the \emph{pairs of
  contexts plugged with $M_1$ and $M_2$}, i.e., tuples such as \(\seq{\Ty,
  \G, \E_0, \Ctx[M_1], \Ctx[M_2], Y}\) where $\Ctx\hole$ is a context
of type $Y$ whose hole expects type $X$. Define the set
$\PluggedContexts{{\Ty_0}\vinf{\G_0}{V_1,V_2}\A}$ for contexts plugged
with values analogously.

For uniformity's sake, we let types $X$ range over both value and
$\E$-computation types, and phrases $P$ range over both value and
computation terms. Judgements of the form $\geninf\Ty\G\E PX$ are
meta-judgements, ranging over value judgements $\Ty\vinf \G PX$ and
$\E$-computation judgement $\Ty\cinf \G P X$.

Let $\Ty;\G \entails_{\E} P,Q \of X$ be two \cbpv{} phrases.  We say
that $P$ and $Q$ are \emph{contextually equivalent} and write
$\geninf\Ty\E\G {P\ceq Q}X$ when, for all pairs of plugged \emph{closed}
\emph{ground-returner} \emph{pure} contexts
\(
\seq{\emptyset, \emptyset, \pure, M_P, M_Q, \F\Gnd}
\) in \(
\PluggedContexts{\geninf{\Ty}{\G}{\E}{P, Q}{X}}
\) and for all closed ground
value terms $\vinf{}V\Gnd$, we have:
\[
  M_P \reducesto^* \return \V
  \qquad \iff \qquad
  M_Q \reducesto^* \return \V
\]

$\cbpv$ has a straightforward set-theoretic denotational
semantics. Presenting the semantics for the core calculus will
simplify our later presentation. To do so, we first recall the
following established facts about monads, specialised and concretised
to the set-theoretic setting.

A monad is a triple $\triple{\T}{\return{}}{\bind{}}$ where $\T$
assigns to each set $X$ a set $\T X$, $\return{}$ assigns to each set
$X$ a function $\return[]{}^X : X \to T X$ and $\bind{}$ assigns to
each function $f : X \to \T Y$ a function $\bind f : \T X \to T Y$, and
moreover these assignments satisfy well-known algebraic identities.
Given a monad $\triple{\T}{\return{}}{\bind{}}$ we define for every
function $f : X \to Y$ the functorial action $\fmap\,f : T X \to T Y$ as
$\fmap\,f\, xs \definedby xs \bind(\return{} \circ f)$.  A
\emph{$T$-algebra} for a monad $\triple{\T}{\return{}}{\bind{}}$ is a
pair $C = \pair{\carrier C}{c_C}$ where $\carrier C$ is a set and
$c_C \of T\carrier C \to \carrier C$ is a function satisfying
\(
  c(\return x) = x
\), and \(
  c(\fmap\ c\ xs) = c(xs \bind \id)
\)
for all $x \in \carrier C$ and $xs \in T^2 \carrier C$.  The set $\carrier C$
is called the \emph{carrier} and we call $c$ the \emph{algebra
  structure}.  For each set $X$, the pair
$F X \definedby \pair{TX}{\bind\id}$ forms a $T$-algebra called the
\emph{free $T$-algebra over $X$}.

\begin{figure}
  \cbpvDenSem
  \caption{\cbpv{} denotational semantics for types}
  \figlabel{cbpv-den-sem}
\end{figure}

We parameterise $\cbpv$'s semantics function $\sem{\kinf \E\EffectKind}$ by
an assignment $\tysem$ of sets $\tysem(\tyv)$ to each of the type
variables $\tyv$ in $\Ty$. Given such a type variable assignment
$\tysem$, we assign to each
\begin{itemize}
\item effect: a monad
  $\ksem{\kinf \E\EffectKind}$, denoted by
  \(
    \triple{\T_{\ksem\E}}{\return{}^{\ksem\E}}{\bind{}^{\ksem\E}}
  \);
\item value type: a set $\ksem{\kinf \A \ValueKind}$;
\item $\E$-computation type: a $\T_{\ksem\E}$-algebra
  $\ksem{\kinf \C{\ComputationKind\E}}$;
and
\item context: the set
  $\ksem{\kinf\G\ContextKind} \definedby \prod_{x \in \Dom\G}\ksem{\G(x)}$.
\end{itemize}

\figref{cbpv-den-sem} defines the standard set-theoretic semantics
function over the structure of types. The pure effect denotes the
identity monad, which sends each set to itself, and extends a function
by doing nothing. The extended languages in the following sections
will assign more sophisticated monads to other effects. The semantics
of type variables uses the type assignment given as parameter. The
unit type always denotes the singleton set. Product types and variants
denote the corresponding set-theoretic operations of cartesian product
and disjoint union, and thus the empty variant type $0 \definedby
\VariantType{}$ denotes the empty set. The type of thunked
$\E$-computations of type $\C$ denotes the carrier of the
$\T_{\ksem\E}$-algebra $\ksem\C$. The $\E$-computation type of $\A$
returners denotes the free $\ksem\E$-algebra. Function and product
types denote well-known algebra structures over the sets of functions
and pairs,
respectively~\citep[Theorem~4.2]{barr:toposes-triples-theories}.

Terms can have multiple types, for example the function $\abst x
{\return x}$ has the types $\vUnitType \to \vUnitType$ and $\SumZero
\to \SumZero$, and type judgements can have multiple type derivations.
We thus give a Church-style semantics~\citep{ReynoldsTheories} by
defining the semantic function for type judgement derivations rather
than for terms.  To increase readability, we write $\sem P$ instead of
including the entire typing derivation for $P$.

The semantic function for terms is parameterised by an
assignment $\tysem$ of sets to type variables. It assigns to each well-typed derivation for a:
\begin{itemize}
\item value term: a function
  $\tsem{\Ty\vinf \G V\A} \of \ksem \G \to \ksem \A$;
  and
\item $\E$-computation term: a function
  $\tsem{\Ty\cinf\G\M\C} \of {\ksem\G\to\carrier{\ksem\C}}$.
\end{itemize}

\figref{cbpv-den-sem-terms} defines the standard set-theoretic
semantics over the structure of derivations. The semantics of
sequencing uses the Kleisli extension function $(\bind f) : TX \to
\carrier{\sem\C}$ for functions into non-free algebras $f : X \to
\carrier{\sem\C}$, given by $(\bind f) \definedby c\compose \return
\compose f$.

\begin{figure}
  \cbpvTermDenSem

  \caption{\cbpv{} denotational semantics for terms}
  \label{fig:cbpv-den-sem-terms}
\end{figure}

\begin{theorem}[\cbpv compositionality]
  The meaning of a term depends only on the meaning of its sub-terms:
  for all pairs of well-typed plugged \cbpv contexts $M_P$, $M_Q$ in
  $\PluggedContexts{{\geninf{\Ty}{\G}{\E}{P, Q}{X}}}$, if $\sem P =
  \sem Q$ then $\sem{M_P} = \sem{M_Q}$.
\end{theorem}

The proof is a straightforward induction on the set of plugged
contexts.

To phrase our simulation results in later development, we adopt a
relaxed variant of simulation: let $\reducestoupto$ be the smallest
relation containing $\breducesto$ that is closed under the term
formation constructs, and so contains $\reducesto$ as well, and let
$\eqcong$ be the smallest congruence relation containing
$\breducesto$.

\begin{theorem}[\cbpv soundness]
  Reduction preserves the semantics: for every pair of well-typed
  \cbpv terms $\geninf \Ty\G\E{P,Q}X$, if $P \eqcong Q$ then $\sem P = \sem
  Q$. In particular, for every well-typed closed term of ground type
  $\geninf {}{}\pure{P}{\F\Gnd}$, if $P \reducesto^* \return V$ then
  $\sem P = \sem V$.
\end{theorem}

The proof is standard: check that $\breducesto$ preserves the
semantics via calculation, and appeal to compositionality.

Combining the \theoremref{cbpv safety}~(safety), \theoremref{cbpv
  termination}~(termination), compositionality, and soundness, we
have:
\begin{theorem}[\cbpv adequacy]
  Denotational equivalence implies contextual equivalence: for all
  well-typed \cbpv terms $\Ty\cinf\G {P,Q} X$, if $\sem P = \sem Q$
  then $P \ceq Q$.
\end{theorem}

As a consequence, we deduce that our operational semantics is very
well-behaved: for all well-typed computations $\Ty\cinf\G{M, M'}\C$,
if $\M \reducestoupto \M'$ then $\M\ceq \M'$.

In the following sections, we will extend the $\cbpv$ calculus using
the following convention. We use an ellipsis to mean that a new
definition consists of the old definition verbatim with the new
description appended, as in the following:
\begin{center}\begin{newsyntax}
  \M, \N &\gdefinedby{}&\cdots  \gor {\operation\op\V} &\synname{effect operation}
\end{newsyntax}
\end{center}

\section{Effect handlers: $\lameff$}\label{sec:eff}
\citet{bauer-pretnar:programming-with-algebraic-effects-and-handlers}
propose algebraic effects and handlers as a basis for modular
programming with user-defined effects.  Programmable effect handlers
arose as part of
\citepos*{plotkin-power:notions-of-computation-determine-monads}{computational
  effects}, which investigates the consequences of using the
additional structure in algebraic presentations of monadic models of
effects. This account refines \citepos*{Moggi89}{ monadic account}
by incorporating into the theory the syntactic constructs that
generate effects as \emph{algebraic operations for a
  monad}~\citep{plotkin-power:algebraic-operations-and-generic-effects}:
each monad is accompanied by a collection of syntactic operations,
whose interaction is specified by a collection of equations, i.e., an
algebraic theory, which fully determines the monad.  To fit exception
handlers into this account,
\citet{plotkin-pretnar:handlers-of-algebraic-effects} generalise to
the handling of arbitrary algebraic effects, giving a computational
interpretation to algebras for a monad. By allowing the user to
declare operations, the user can describe new effects in a composable
manner. By defining algebras for the free monad with these operations,
users give the abstract operations different meanings similarly to
\citepos*{Swierstra08}{use of free monads}.

\begin{figure}
\begin{tabular}{c@{\ \ }c}
  \lameffSyntax{}
&
  \lameffOpSemantics{}
  \\\\
  \subcaptionn{Syntax extensions to \figref{cbpv-syntax}}
             {lameff-syntax}
  &
  \subcaptionn{Operational semantics extensions to \figref{cbpv-op-semantics}}
             {lameff-op-semantics}
\end{tabular}
  \caption{$\lameff$}\figlabel{lameff}
\end{figure}

\subfigref{lameff}{lameff-syntax} presents the extension $\lameff$,
Kammar et al.'s core calculus of effect
handlers~\citep{kammar-lindley-oury:handlers-in-action}. We assume a
countable set of elements of a separate syntactic class, ranged over
by $\op$. We call these \emph{operation names}. For each operation
name $\op$, $\lameff$'s operation call construct allows the programmer
to invoke the effect associated with $\op$ by passing it a value as an
argument. Operation names are the only interface to effects the
language has. The handling construct allows the programmer to use a
handler to interpret the operation calls of a given returner
computation. As the given computation may call thunks returned by
functions, the decision which handler will handle a given operation
call is dynamic. Handlers are specified by two kinds of clauses. A
\emph{return clause} describes how to proceed when returning a
value. An \emph{operation clause} describes how to proceed when
invoking an operation $\op$. The body of an operation clause can
access the value passed in the operation call using the first bound
variable $p$, which is similar to the bounding occurrence of an
exception variable when handling exceptions. But unlike exceptions, we
expect arbitrary effects like reading from or writing to memory to
resume. Therefore the body of an operation clause can also access the
continuation at the operation's calling point. Even though we use a
list notation in this presentation of the syntax, the abstract syntax
tree representation of a handler $\H$ is in fact a pair $H =
\pair{\handlerRet\H}{\handlerOp\H\placeholder}$ consisting of a single
return clause $\handlerRet\H$, and a function
$\handlerOp\H\placeholder$ from a finite subset of the operation names
assigning to each operation name $\op$ its associated operation clause
$\handlerOp\H\op$.

\example The two left columns of \figref{state eff} demonstrate how to add
user-defined boolean state in $\lameff$. The handler $\H_{ST}$ is
parameterised by the current state. When the computation terminates,
we discard this state. When the program calls $\op[get]$, the handler
returns the current state and leaves it unchanged. When the program
calls $\op[put]$, the handler returns the unit value, and instates the
newly given state.

\begin{figure}
  \def\SharedSpace{.9em}
  \begin{tabular}{c@{\hspace{\SharedSpace}}c@{\hspace{\SharedSpace}}c}
  \(
    \var{toggle} = \thunk{
      \begin{array}[t]{@{}l}
      \Let{x}{\operation{\op[get]}{\vunit}}\\
      \Let{y}{\apply{\force{\var{not}}}{x}}\\
      \Let*{\operation{\op[put]}{y}}\\
       x
    }
      \end{array}
      \)
      &
      \(
      \begin{array}[t]{@{}l@{}l@{}}
      \H_{ST} &{}= \begin{array}[t]{@{}l@{}l}
        \{
               \begin{array}[t]{@{}l@{\,}l@{~}c@{~}l@{}l@{~}l@{}}
               \multicolumn{2}{@{}l@{}}{\return~x}     &\mapsto& \abst{s&}{\mathrlap{\return x}} \\
               \op[get]~\_ & k &\mapsto& \abst{s&}{\force{k}~s~&s} \\
               \op[put]~s' &k  &\mapsto& \abst{\_&}{\force{k}~()&s'}\} \\
               \end{array}
      \end{array}
      \\
        \var{runState} &{}=
    \thunk{\abst{c}
      {       \handle{\force c}
              {\H_{ST}}}}
      \end{array}
      \)
      &
      \(
\begin{array}[t]{@{}l@{}}
\var{State} = \set{\op[get] : \Fun\vUnitType\bit, \op[put] : \Fun\bit\vUnitType} : \EffectKind
\\
\var{toggle} : \U{\var{State}}\F\bit
\\
\H_{ST} : \HandlerType \bit{\var{State}}{\Fun\bit{\F\bit }}\pure
\\
\var{runState} : \U\pure(\Fun{(\U{\var{State}}\F\bit)}{\Fun\bit{\F\bit}})
\end{array}
\)
  \end{tabular}
  \caption{User-defined boolean state in \lameff}\figlabel{state eff}
\end{figure}

\subfigref{lameff}{lameff-op-semantics} presents $\lameff$'s extension
to $\cbpv$'s operational semantics. Computation frames $\EF$ now
include the handling construct, while the basic frames $\EB$ do not,
allowing a handled computation to $\beta$-reduce under the handler. We
add two $\beta$-reduction cases. When the returner computation inside
a handler is fully evaluated, the return clause proceeds with the
return value.  When the returner computation inside a handler needs to
evaluate an operation call, the definition of hoisting contexts $\CF$
ensures $\CF$ is precisely the continuation of the operation call
delimited by the handler. Put differently, it ensures that the handler
in the root of the reduct is the closest handler to the operation call
in the call stack.  The operation clause corresponding to the
operation called then proceeds with the supplied parameter and current
continuation. Rewrapping the handler around this continuation ensures
that all operation calls invoked in the continuation are handled in
the same way. An
alternative~\citep{\hia,KiselyovSS13,lindley-mcbride-mclaughlin:dobedobedo}
is to define instead:
\[
  \handle{\CF[{\operation{\op}\V}]}{\H} 
  \breducesto 
  \substitute{\N}{ \for{\V}{\p}, \for{\thunk{\abst\x{\CF[\return
          \x]}}}{\k}}
\] This variant is known as \emph{shallow} handlers, as opposed to the
\emph{deep} handlers of \subfigref{lameff}{lameff-op-semantics}. We
focus on deep handlers as they are closer to monadic reflection and
have a clean denotational semantics.

\example With this semantics, the user-defined state from \figref{state eff} behaves as expected: \[
\force{\var{runState}}~\var{toggle}~\tru \reducesto^*
\apply{(\handle \tru {\H_{ST}})} \fls
\reducesto^*
\tru
\]
More generally, the handler $\H_{ST}$ expresses \emph{dynamically
  scoped}
state~\cite{kammar-pretnar:no-value-restriction-is-needed-for-algebraic-effects-and-handlers}.
For additional handlers for state and other effects, see
\citepos{pretnar:tutorial}{ tutorial}.

\begin{figure}
  \lameffKindingAndTypingExtension
  \caption{$\lameff$'s kinding and typing (extending
    \figmul~\figref*{cbpv-types}
    and~\figref*{cbpv-kind-and-type-system})}
    \figlabel{lameff-types-syntax}
    \figlabel{lameff-types-and-kinds}
\end{figure}

\figref{lameff-types-syntax} presents $\lameff$'s extension to the
kind and type system.  The effect annotations in $\lameff$ are
functions from finite sets of operation names, assigning to each
operation name its parameter type $A$ and its return type $B$. We add
a new kind for handler types, which describe the kind and the returner
type the handler can handle, and the kind and computation type the
handling clause will have.

In the kinding judgement for effects, the types in each operation's
arity assignment must be value types. The kinding judgement for
handlers requires all the types and effects involved to be
well-kinded.
Computation type judgements now include two additional rules for each
new computation construct. An operation call is well-typed when the
parameter and return type agree with the arity assignment in the
effect annotation. A use of the handling construct is well-typed when
the type and effect of the handled computation and the type-and-effect
of the construct agree with the types and effects in the handler
type. The set of handled operations must strictly agree with the set
of operations in the effect annotation. The variable bound to the
return value has the returner type in the handler type. In each
operation clause, the bound parameter variable has the parameter type
from the arity assignment for this operation, and the continuation
variable's input type matches the return type in the operation's arity
assignment. The overall type of all operation clauses agrees with the
computation type of the handler. The second effect annotation on the
handler type matches the effect annotations on the continuation and
the body of the operation and return clauses, in accordance with the
deep handler semantics.

\example The type system assigns the boolean state terms the types given in
\figref{state eff}.

$\lameff$'s design involves several decisions. First, handlers have
their own kind, unlike Pretnar's calculus in which they are
values~\citep{pretnar:tutorial}. This distinction is minor, as
handlers as values can be expressed by thunking the handling
construct, cf.~$\H_{ST}$ and $\var{runState}$ above. Next, the effect
annotations involved in the handling construct have to agree
precisely. Another option is to check inclusion of operation sets,
i.e., a handler may handle more effects than the annotation on the
effect. This distinction is minor, as we can express coercions from an
effect annotation into a superset of effects using a trivial handler:
\[
  \set{\abst \x{\return \x}}\uplus\set{\handlerEntry{\op}{\p}{\k}{\k
      (\operation\op \p)}\suchthat \op\in\E} \of \HandlerType{\A}{\E}{\F\A}{\E\uplus\E'}
\]
A more significant choice is to use \emph{closed} handlers: execution
halts/crashes when a handled computation calls an operation the
handler does not handle. The other option is to use \emph{forwarding}
handlers~\citep{\hia}, in which unhandled operation calls are forwarded
to the nearest enclosing handler that can handle them. In our simple
type-and-effect system, this decision has no immediate impact, as we
can use the trivial handler above to re-raise unhandled effects whenever
needed. However, in more expressive type systems, which we do not
consider here, in particular type systems with \emph{effect
  polymorphism}~\citep{
  lucassen-gifford:polymorphic-effect-systems,
  leijen:koka-handlers,
  KiselyovSS13,
  lindley-hillerstrom:liberating-effects-with-rows-and-handlers
},
this distinction is more significant. In this case, we believe that
the language should include both variants: the forwarding variant to
support code extensibility and modularity, and the closed variant to
allow the programmer to guarantee that a computation cannot cause
unhandled effects, or a mechanism for ascribing effect annotations to
ensure all effects have been handled. Finally, it is possible to
remove the effect system. In that case, the arity assignments for the
operations need to be placed globally at the top level of the program,
as in Pretnar's tutorial~\citep{pretnar:tutorial}. Removing the effect
system has dramatic consequences on expressivity: as we are about to
see, well-typed $\lameff$ terms always terminate. If we remove the
effect annotations, we can encode a form of Landin's
knot~\citep{landin:the-mechanical-evaluation-of-expressions}, making
the calculus non-terminating.

\lameff's meta-theoretic development follows \cbpv's development
closely, with an Abella formalisation of safety:
\begin{theorem}[\lameff safety]\theoremlabel{eff safety}
  Well-typed programs don't go wrong: for all closed \lameff returners
  $\Ty\einf\emptyset{}M{\F\A}$, either $M \reducesto N$ for some
  $\Ty\einf\emptyset{}{N}{\F\A}$ or else ${M = \return V}$ for some
  $\Ty\vinf {}V\A$.
\end{theorem}

Using the monadic lifting from \citepos*{kammar:thesis}{thesis}, we
obtain termination for $\lameff$~\cite{\hia}:
\begin{theorem}[\lameff termination]\theoremlabel{eff termination}
  There are no infinite reduction sequences: for all \lameff terms
  $\cinf[\pure]{}M{\F\A}$, we have $M \not\diverges$, and there exists
  some unique $\vinf {}\V\A$ such that
  \( M \reducesto^\star \return \V \).
\end{theorem}

$\lameff$ shares $\cbpv$'s ground types, and we define plugged
contexts and the equivalences $\ceq$ and $\eqcong$ as in $\cbpv$.

We give an adequate set-theoretic denotational semantics for
$\lameff$. First, recall the following well established concepts in
universal and categorical algebra.
A \emph{signature} $\sig$ is a pair consisting of a set $\carrier
\sig$ whose elements we call \emph{operation symbols}, and a function
$\arity[\sig]$ from $\carrier\sig$ assigning to each operation symbol
$f \in \carrier\sig$ a (possibly infinite) set $\arity(f)$. We write
$(f : A) \in \sig$ when $f \in \carrier\sig$ and ${\arity[\sig] (f) =
  A}$. Given a signature $\sig$ and a set $X$, we inductively form the
set $\T_{\sig}X$ of \emph{$\sig$-terms over $X$} by:
\[
  t  \gdefinedby
         x
    \gor f\seq[a \in A]{t_a} \hphantom{foobar}\qquad (x \in X, (f : A) \in \sig)
\]
The assignment $\T_{\sig}$ together with the following assignments form a monad
\begin{mathpar}
  \return x \definedby x

t \bind f \definedby \substitute t{\for{f(x)}x}_{x \in X}\quad(f \of X \to \T_{\sig} Y)
\end{mathpar}
The $\T_{\sig}$-algebras $\pair Cc$ are in bijective correspondence
with \emph{$\sig$-algebras} on the same carrier. These are pairs
$\pair C{\sem\placeholder}$ where $\sem{\placeholder}$ assigns to each
$(f : A) \in \sig$ a function $\sem\placeholder : C^A \to C$ from
$A$-ary tuples of $C$ elements to $C$. The bijection is given by setting
\(
\sem f\seq[a \in A]{\xi_a}
\)
to be
\(
c(f\seq[a \in A]{\xi_a})
\).

$\lameff$'s denotational semantics is given by extending $\cbpv$'s semantics as
follows. Given a type variable assignment $\tysem$, we assign to each
\makeatletter
\begin{itemize}
\item[\llap{$\cdots\quad$}\@itemlabel] 
  handler type: a pair $\sem{\kinf X\HandlerKind} = \pair Cf$
  consisting of an algebra $C$ and a function $f$ into the $\carrier
  C$ carrier of this algebra.
\end{itemize}
  \makeatother

\figref{lameff-den-sem} presents how $\lameff$ extends $\cbpv$'s
denotational semantics. Each effect $\E$ gives rise to a
signature whose operation symbols are the operation names in $\E$
tagged by an element of the denotation of the corresponding parameter
type. This signature gives rise to the monad $\E$ denotes. When $\E =
\pure$, the induced signature is empty, and gives rise to the identity
monad, and so this semantic function extends $\cbpv$'s
semantics. Handlers handling $\E$-computations returning $\A$-values
using $\E'$-computations of type $\C$ denote a pair. Its first
component is an $\ksem\E$-algebra structure over the carrier
$\carrier{\ksem\C}$, which may have nothing to do with the
$\ksem{\E'}$-algebra structure $\ksem\C$ already possesses. The second
component is a function from $\ksem\A$ to the carrier
$\carrier{\ksem\C}$.

The denotation of an operation call to $\op$ makes use of the fact that
the effect annotation $\E$ contains the operation name
$\op$. Consequently, the resulting signature contains an operation
symbol $\op_q$ for every $q \in \ksem\Pa$. The denotation of $\op$ is
then the term $\op_q\seq[a \in \ksem\Ar]a$. The denotation of the
handling construct uses the Kleisli extension of the second component
in the denotation of the handler. The denotation of a handler term
defines the $\T_{\sig}$-algebras by defining a $\sig$-algebra for the
associated signature $\sig$. The operation clause for $\op$ allows us
to interpret each of the operation symbols associated to $\op$. The
denotation of the return clause gives the second component of the
handler.

\begin{figure}
\def\SharedSpace{3pt}
\begin{tabular}{@{}l@{\hspace{.25cm}} l@{}}
  \lameffDenSem{}\medskip\\
  \lameffTermDenSem
  \end{tabular}
  \caption{\lameff denotational semantics (extending
    \figmul~\figref*{cbpv-den-sem}
    and~\figref*{cbpv-den-sem-terms})}
  \figlabel{lameff-den-sem-types}
  \figlabel{lameff-den-sem-terms}
  \figlabel{lameff-den-sem}
\end{figure}

\begin{theorem}[\lameff compositionality]
  The meaning of a term depends only on the meaning of its sub-terms:
  for all pairs of well-typed plugged \lameff contexts $M_P$, $M_Q$ in
  $\PluggedContexts{{\geninf{\Ty}{\G}{\E}{P, Q}{X}}}$, if $\sem P =
  \sem Q$ then $\sem{M_P} = \sem{M_Q}$.
\end{theorem}

The proof is identical to $\cbpv$, with two more cases for
$\breducesto$. Similarly, we have:

\begin{theorem}[\lameff soundness]
  Reduction preserves the semantics: for every pair of well-typed
  \lameff terms $\geninf \Ty\G\E{P,Q}X$, if $P \eqcong Q$ then $\sem P = \sem
  Q$. In particular, for every well-typed closed term of ground type
  $\geninf {}{}\pure{P}{\F\Gnd}$, if $P \reducesto^* \return V$ then
  $\sem P = \sem V$.
\end{theorem}

We combine the previous results, as with \cbpv:

\begin{theorem}[\lameff adequacy]
  Denotational equivalence implies contextual equivalence: for all
  well-typed \lameff terms $\Ty\cinf\G {P,Q} X$, if $\sem P = \sem Q$
  then $P \ceq Q$.
\end{theorem}

Therefore, \lameff also has a well-behaved operational semantics: for
all well-typed computations $\Ty\cinf\G{M, M'}\C$, if $\M
\reducestoupto \M'$ then $\M\ceq \M'$.

\section{Monadic reflection: $\lammon$}\label{sec:mon}

Languages that use monads as an abstraction for user-defined effects
employ other mechanisms to support them, usually an overloading
resolution mechanism, such as type-classes in Haskell and Coq, and
functors/implicits in OCaml. As a consequence, such accounts for
monads do not study them as an abstraction in their own right, and are
intertwined with implementation details and concepts stemming from the
added mechanism.  Filinski's work on monadic
reflection~\citep{filinski:representing-monads,
                  filinski:thesis,
                  filinski:representing-layered-monads,
                  filinski:monads-in-action}
serves precisely this purpose: a calculus in which user-defined monads
stand independently.

\begin{figure}[b]
  \begin{tabular}{cc}
    \lammonSyntax{}
    &
      \lammonOpSemantics{}
    \\\\
    \subcaptionn{Syntax extensions to \figref{cbpv-syntax}}
               {lammon-syntax}
    &
    \subcaptionn{Operational semantics extensions to \figref{cbpv-op-semantics}}
               {lammon-op-semantics}
  \end{tabular}
  \caption{$\lammon$}
  \figlabel{lammon}
\end{figure}

\subfigref{lammon}{lammon-syntax} presents $\lammon$'s syntax. The \(
\effdef* \x{N_u}\y f{N_b} \) construct binds $\x$ in the term $N_u$
and $\y$ and $f$ in $N_b$. The term $N_u$ describes the unit and the
term $N_b$ describes the Kleisli extension/bind operation. We
elaborate on the choice of the keyword $\keyword{where}$ when we
describe $\lammon$'s type system.  Using monads, the programmer can
write programs as if the new effect was native to the language. We
call the mode of programming when the effect appears native the
\emph{opaque} view of the effect. In contrast, the \emph{transparent}
mode occurs when the code can access the implementation of the effect
directly in terms of its defined monad. The \emph{reflect} construct
$\reflect N$ allows the programmer to graft code executing in
transparent mode into a block of code executing in opaque mode. The
\emph{reify} construct $\reify N$ turns a block of opaque code into
the result obtained by the implementation of the effect.

\example \figref{state mon} demonstrates how to add user-defined boolean state
in \lammon using the standard $\var{State}$ monad.  To express
$\var{get}$ and $\var{put}$, we reflect the concrete definition of the
corresponding operations of the state monad.  To run a computation, we
use reification to get the monadic representation of the computation
as a state transformer, and apply it to the initial state.

\begin{figure}
  \def\SharedSpace{1em}
  \begin{tabular}{@{}c@{\hspace{\SharedSpace}}c@{\hspace{\SharedSpace}}c@{}}
    \(
    \begin{array}[t]{@{}l@{}l@{}l@{}l@{}}
      \var{toggle} =  \mathrlap{\thunk{
        \begin{array}[t]{@{}l@{}}
          \Let{x}{\force{\var{get}}}\\
          \Let{y}{\apply{\force{\var{not}}}{x}}\\
          \Let*{\force{\var{put}}~y}\\
          x
      }
        \end{array}}
      \\
      \var{get} = \thunk{&\reflect{\abst{s}{(s&, s&)}}} \\
      \var{put} = \thunk{\abst{s'}{&\reflect{\abst{\_}{(()&, s'&)}}}} \\
    \end{array}
    \)
    &
    \(
    \begin{array}[t]{@{}l@{}l@{}}
      \var{State} &{}=\\
      \multicolumn{2}{@{}c@{}}{
        \begin{array}[t]{@{\qquad}l@{\,}l@{~}c@{~}l@{}l@{~}l@{}}
          \mspace{-18mu}\effdef*[\\]{x&}{\abst{s}{{\vpair xs}}}
                            {\\f}{k&}{\abst{s}{\begin{array}[t]{@{}l@{}}
                                  \Let{(x,s')}{f~s}{\\
                                    \force{k}~x~s'}}}
                                \end{array}
        \end{array}
      }
      \\
      \var{runState} &{}=
      \thunk{
        \abst{c}
             {\reify
               [\var{State}]
               {\force{c}}}}
    \end{array}
    \)
    &
    \(
    \begin{array}[t]{@{}l@{}}
      \effType*[{\\\qquad}] \pure\alpha{\Fun\bit\F(\vProdType\alpha\bit)}{\var{State}} : \EffectKind
      \\
      \var{toggle} : \U{\var{State}}\F\bit
      \\
      \var{runState} : \U\pure(\Fun{(\U{\var{State}}\F\bit)}{\Fun\bit{\F(\vProdType\bit\bit)}})
      \\
      \var{get} : U_{\var{State}}\F\bit
      \\
      \var{put} : U_{\var{State}}{(\Fun\bit\F\vUnitType)}
\end{array}
\)
  \end{tabular}
  \caption{User-defined boolean state in \lammon}\figlabel{state mon}
\end{figure}

\subfigref{lammon}{lammon-op-semantics} describes the extension to the
operational semantics. The $\reifyRule$ transition uses the
user-defined monadic return to reify a value. To explain the
$\reflectRule$ transition, note that the hoisting context $\CF$
captures the continuation at the point of reflection, with an opaque
view of the effect $\Monad$. The reflected computation $N$ views this
effect transparently. By reifying $\CF$, we can use the user-defined
monadic bind to graft the two together.

\example With this semantics we have
\(
\force{\var{runState}}~\var{toggle}~\tru \reducesto^\star \return~(\tru, \fls)
\).

The example we have given here fits with the way in which monadic
reflection is often used, but is not as flexible as the effect handler
version because $\var{get}$ and $\var{put}$ are concrete functions
rather than abstract operations, which means we cannot abstract over
how to interpret them. To write a version of toggle that can be
interpreted in different ways is possible using monadic reflection but
requires more sophistication. We shall see how to do so once we have
defined the translation of $\lameff$ into $\lammon$.

\begin{figure}[b]
  \lammonKindingAndTypingExtension
  \caption{$\lammon$'s kinding and typing (extending
    \figmul~\figref*{cbpv-types}
    and~\figref*{cbpv-kind-and-type-system})}
    \figlabel{lammon-types}
    \figlabel{lammon-type-and-kind-system}
\end{figure}

\figref{lammon-type-and-kind-system} presents the natural extension to
$\cbpv{}$'s kind and type system for monadic reflection. Effects are a
stack of monads. The empty effect is the identity monad. A monad
$\Monad$ can be \emph{layered} on top of an existing stack $\E$:
\[
\effType \E\alpha C{\effdef*\x\M\y f\N}
\]
The intention is that the type constructor
$\substitute\C{\for\placeholder\alpha}$ has an associated monad
structure given by the bodies of the return $M$ and the bind $N$, and
can use effects from the rest of the stack $\E$. To be well-kinded,
$C$ must be an $\E$-computation, and $\Monad$ must be a well-typed
monad, i.e., the return should have type $\substitute C{\for A\alpha}$
when substituted for some value $V : A$, and the bind should implement
a Kleisli extension operation.

\example \figref{state mon} demonstrates a kind and type assignment to the user-defined
global state example.

The choice of keywords for monads and their types is modelled on their
syntax in Haskell. We stress that our calculus does not, however,
include a type-class mechanism. The \emph{type} of a monad contains
the return and bind \emph{terms}, which means that we need to check
for equality of terms during type-checking, for example, to ensure
that we are sequencing two computations with compatible effect
annotation. For our purposes, $\alpha$-equivalence suffices. This need
comes from our choice to use structural, anonymous, monads. In
practice, monads are given \emph{nominally}, and two monads are
compatible if they have exactly the same name. It is for this reason
also that the bodies of the return and the bind operations must be
closed, apart from their immediate arguments. If they were allowed to
contain open terms, types in type contexts would contain these open
terms through the effect annotations in thunks, requiring us to
support dependently-typed contexts. The monad abstraction is
parametric, so naturally requires the use of type variables, and for
this reason we include type variables in the base calculus $\cbpv$. We
choose monads to be structural and closed to keep them closer to the
other abstractions and to reduce the additional lingual constructs
involved.

Our calculus deviates from
\citepos{filinski:monads-in-action} in the following
ways. First, our effect definitions are local and structural, whereas his
allow nominal declaration of new effects only at the top
level. Because we do not allow the bodies of the return and the bind to
contain open terms, this distinction between the two calculi is
minor. As a consequence, effect definitions in both calculi are
\emph{static}, and the monadic bindings can be resolved at compile
time. Filinski's calculus also includes a sophisticated
\emph{effect-basing} mechanism, that allows a computation to
immediately use, via reflection, effects from any layer in the
hierarchy below it, whereas our calculus only allows reflecting
effects from the layer immediately below.
In the presence of Filinski's type system, this deviation does not
significantly change the expressiveness of the calculus: the monad
stack is statically known, and, having access to the type information,
we can insert multiple reflection operators and lift effects from
lower levels into the current level.

We also prove \lammon's Felleisen-Wright soundness in our Abella
formalisation:
\begin{theorem}[\lammon safety]\theoremlabel{mon safety}
  Well-typed programs don't go wrong: for all closed \lammon returners
  $\Ty\einf\emptyset{}M{\F\A}$, either $M \reducesto N$ for some
  $\Ty\einf\emptyset{}{N}{\F\A}$ or else ${M = \return V}$ for some
  $\Ty\vinf {}V\A$.
\end{theorem}

As with $\lameff$, $\lammon$'s ground types are the same as $\cbpv$'s.
While we can define an observational equivalence relation in the same
way as for $\cbpv{}$ and $\lameff$, we will not do so.  Monads as a
programming abstraction have a well-known conceptual complication ---
user-defined monads must obey the \emph{monad laws}. These laws are a
syntactic counterpart to the three equations in the definition of
(set-theoretic/categorical) monads. The difficulty involves deciding
what equality between such terms means. The natural candidate is
observational equivalence, but as the contexts can themselves define
additional monads, it is not straightforward to do so. Giving an
acceptable operational interpretation to the monad laws is an open
problem. We avoid the issue by giving a \emph{partial} denotational
semantics to $\lammon$.

\begin{figure}
  \begin{tabular}[t]{@{}l@{\qquad}l@{}}
    \begin{tabular}[t]{@{}l@{}}
      \lammonDenSem{}
    \end{tabular}
    &
    \begin{tabular}[t]{@{}l@{}}
      \lammonDenSemTerms
    \end{tabular}
  \end{tabular}
  \vspace{-4ex}
  \caption{\lammon denotational semantics (extending
    \figmul~\figref*{cbpv-den-sem}
    and~\figref*{cbpv-den-sem-terms})}
  \figlabel{lammon-den-sem}
  \figlabel{lammon-den-sem-types}
  \figlabel{lammon-den-sem-terms}
\end{figure}

Extend $\cbpv{}$'s denotational semantics to $\lammon$ as follows. Given a type variable assignment $\tysem$, we assign to each
\makeatletter
\begin{itemize}
\item[\llap{$\cdots\quad$}\@itemlabel] 
  monad type and effect: a monad $\sem{\Ty\minf \Monad \E}\tysem = \sem{\kinf\E\EffectKind}\tysem$, if the sub-derivations
  have well-defined denotations, and this data does indeed form a
  set-theoretic monad.
\end{itemize}
\makeatother
Consequently, the denotation of any derivation is undefined if at
least one of its sub-derivations has undefined semantics. Moreover,
the definition of kinding judgement denotations now depend on term
denotation.

\figref{lammon-den-sem} shows how $\lammon$ extends $\cbpv$'s
denotational semantics. The assigned type-constructor, and
user-defined return and bind, if well-defined, have the appropriate
type to give the structure of a monad, and the semantics's definition
posits they do. For the term semantics, recall that \(
\Monad_{\sem{\effType \E\alpha C\Monad}}X =
\carrier{\tsem[{(\theta[\alpha \mapsto X])}]{C}} \) and therefore,
semantically, we can view any computation of type and kind
\(
\kinf {\F \A} {\ComputationKind{\effType \E\alpha C\Monad}}
\)
as an $\E$-computation of type $\substitute C{\for A\alpha}$.

We define a \emph{proper derivation} to be a derivation whose
semantics is well-defined for all type variable assignments, and a
\emph{proper term or type} to be a term or type that has a proper
derivation.  Thus, a term is proper when all the syntactic monads it
contains denote semantic set-theoretic monads. When dealing with the
typed fragment of \lammon, we restrict our attention to such proper
terms as they reflect the intended meaning of monads. Doing so allows
us to mirror the meta-theory of \cbpv and \lameff for proper terms.

We define \emph{plugged proper contexts} as with \cbpv and \lameff
with the additional requirement that all terms are proper. The
definitions of the equivalences $\ceq$ and $\eqcong$ are then
identical to those of \cbpv and \lameff.

\begin{theorem}[\lammon termination]\theoremlabel{mon termination}
  There are no infinite reduction sequences: for all \emph{proper}
  \lammon terms ${\cinf[\pure]{}M{\F\A}}$, we have $M \not\diverges$,
  and there exists some unique $\vinf {}\V\A$ such that \( M
  \reducesto^\star \return \V \).
\end{theorem}

Our proof uses
\citepos*{lindley-stark:reducibility-and-TT-lifting-for-computation-types}{$\top\top$-lifting}.

\begin{theorem}[\lammon compositionality]
  The semantics depends only on the semantics of sub-terms:
  for all pairs of well-typed plugged \emph{proper} \lammon contexts
  $M_P$, $M_Q$ in $\PluggedContexts{{\geninf{\Ty}{\G}{\E}{P, Q}{X}}}$,
  if $\sem P = \sem Q$ then $\sem{M_P} = \sem{M_Q}$.
\end{theorem}

The proof is identical to $\cbpv$, with two more cases for
$\breducesto$. Similarly, we have:

\begin{theorem}[\lammon soundness]
  Reduction preserves the semantics: for every pair of well-typed
  \emph{proper} \lammon terms $\geninf \Ty\G\E{P,Q}X$, if $P \eqcong
  Q$ then $\sem P = \sem Q$. in particular, for every well-typed
  closed term of ground type $\geninf {}{}\pure{P}{\F\Gnd}$, if $P
  \reducesto^* \return V$ then $\sem P = \sem V$.
\end{theorem}

We combine the previous results, as with \cbpv and \lameff:

\begin{theorem}[\lammon adequacy]
  Denotational equivalence implies contextual equivalence: for all
  well-typed \emph{proper} \lammon terms $\Ty\cinf\G {P,Q} X$, if
  $\sem P = \sem Q$ then $P \ceq Q$.
\end{theorem}

Therefore, the \emph{proper} fragment of \lammon also has a
well-behaved operational semantics: for all well-typed proper
computations $\Ty\cinf\G{M, M'}\C$, if $\M \reducestoupto \M'$ then
$\M\ceq \M'$.

In contrast to $\lameff$ the semantics for $\lammon$ is finite:
\begin{lemma}[finite denotation property]\lemmalabel{finite model lammon}
  For every type variable assignment $\theta = \seq[\alpha \in
    \Theta]{X_\alpha}$ of \emph{finite} sets, every proper \lammon
  value type $\kinf\A$ and computation type $\kinf \C$ denote finite
  sets $\tsem{\A}$, $\tsem{\C}$.
\end{lemma}

\section{Delimited control: $\lamdel$}\label{sec:del}
Delimited control operators can implement algorithms with
sophisticated control structure, such as tree-fringe comparison, and
other control mechanisms, such as coroutines
\citep{felleisen:the-theory-and-practice-of-first-class-prompts} yet
enjoy an improved meta-theory in comparison to their undelimited
counterparts~\citep{felleisn-wand-friedman-duba:abstract-continuations}.
The operator closest in spirit to handlers, $\keyword{S_0}$
pronounced ``shift zero'', was introduced by
Danvy and Filinski~\cite{danvy-filinski:abstracting-control} as part of a systematic
study of continuation-passing-style conversion.

\begin{figure}
\begin{tabular}{cc}
  \lamdelSyntax{}
  &
  \lamdelOpSemantics{}
  \\\\
  \subcaptionn{Syntax extensions to \figref{cbpv-syntax}}
              {lamdel-syntax}
  &
  \subcaptionn{Operational semantics extensions to \figref{cbpv-op-semantics}}
              {lamdel-op-semantics}
\end{tabular}
  \caption{$\lamdel$}
  \figlabel{lamdel}
\end{figure}

\subfigref{lamdel}{lamdel-syntax} presents the extension
$\lamdel$. The construct $\ShiftZ kM$, which we often call ``shift''
(as we find ``shift zero'' awkward), captures the current continuation
and binds it to $k$, and replaces it with $M$. The construct $\Reset
MxN$, which we will call ``reset'', delimits any continuations
captured by shift inside $M$. Once $M$ runs its course and returns a
value, this value is bound to $x$ and $N$ executes. For delimited
control cognoscenti this construct is known as ``dollar'', and it is
capable of macro expressing the entire CPS
hierarchy~\citep{materzok-biernacki:a-dynamic-interpretation-of-the-CPS-hierarchy}.

\example \figref{state del} demonstrates how to add user-defined boolean state
in \lamdel~\cite{danvy-advanced-thesis}. The code assumes the
environment outside the closest reset will apply it to the currently
stored state. By shifting and abstracting over this state, $\var{get}$
and $\var{put}$ can access this state and return the appropriate
result to the continuation. When running a stateful computation, we
discard the state when we reach the final return value.

\begin{figure}[b]
  \def\SharedSpace{3em}
  \begin{tabular}{@{}c@{\hspace{\SharedSpace}}c@{\hspace{\SharedSpace}}c@{}}
    \(
      \var{toggle} =  \thunk{
        \begin{array}[t]{@{}l@{}}
          \Let{x}{\force{\var{get}}}\\
          \Let{y}{\apply{\force{\var{not}}}{x}}\\
          \Let*{\force{\var{put}}~y}\\
          x
      }
        \end{array}
    \)
    &
    \(
    \begin{array}[t]{@{}l@{}l@{}l@{}l@{}l@{}}
      \var{get} &{}= \thunk{&\ShiftZ{k}{\abst{s}{\force{k}~&s~s&}}}\\
      \var{put} &{}= \thunk{\abst{s'}{&\ShiftZ{k}{\abst{\_}{\force{k}~{}(&)~s'&}}}}\\
      \var{runState} &
      \multicolumn{3}{@{}l@{}}{{}=\thunk{\abst{c}{\Reset{\force c}{x}{\abst{s}{x}}}}}
    \end{array}
    \)
    &
    \(
    \begin{array}[t]{@{}l@{}}
      \var{State} = \pure,\Fun\bit{\F\bit} : \EffectKind
      \\
      \var{toggle} : \U{\var{State}}\F\bit
      \\
      \var{runState} : \U\pure(\Fun{(\U{\var{State}}\F\bit)}{\Fun\bit{\F\bit}})
      \\
      \var{get} : U_{\var{State}}\F\bit
      \\
      \var{put} : U_{\var{State}}{(\Fun\bit\F\vUnitType)}
\end{array}
\)
  \end{tabular}
  \caption{User-defined boolean state in \lamdel}\figlabel{state del}
\end{figure}

The extension to the operational semantics in
\subfigref{lamdel}{lamdel-op-semantics} reflects our informal
description. The $\resetRule$ rule states that once the delimited
computation returns a value, this value is substituted in the
remainder of the reset computation. For the $\shiftRule$ rule, the
definition of hoisting contexts guarantees that in the reduct
$\Reset{\CF[{\ShiftZ kM}]}xN$ there are no intervening resets in
$\CF$, and as a consequence $\CF$ \emph{is} the delimited continuation
of the evaluated shift. After the reduction takes place, the
continuation is re-wrapped with the reset, while the body of the shift
has access to the enclosing continuation. If we were to, instead, not
re-wrap the continuation with a reset, we would obtain the
control/prompt-zero operators,
(cf.~\citepos{shan:a-static-simulation-of-dynamic-delimited-control}
and
\citepos{kiselyov-friedman-sabry:static-and-dynamic-delimited-continuation-operators-are-equally-expressible}
analyses of macro expressivity relationships between these two, and
other, variations on untyped delimited control).

\example With this semantics, we have:
\[
\force{\var{runState}}~\var{toggle}~\tru
\reducesto^*
\apply{\Reset{\tru}{x}{\abst s{x}}}{\fls}
\reducesto^*
\return~\tru
\]

\begin{figure}
  \lamdelKindingAndTypingExtension
  \caption{$\lamdel$'s kinding and typing (extending
    \figmul~\figref*{cbpv-types}
    and~\figref*{cbpv-kind-and-type-system})}
    \figlabel{lamdel-types}
    \figlabel{lamdel-types-and-kinds}
\end{figure}

\figref{lamdel-types-and-kinds} presents the natural extension to
$\cbpv{}$'s kind and type system for delimited control. It is based on
Danvy and Filinski's
description~\citep{danvy-filinski:a-functional-of-typed-contexts};
they were the first to propose a type system for delimited control.
Effects are now a stack of computation types, with the empty effect
standing for the empty stack. The top of this stack is the return type
of the currently delimited continuation. Thus, as
\figref{lamdel-types-and-kinds} presents, a
shift pops the top-most type off this stack and uses it to type the
current continuation, and a reset pushes the type of the
delimited return typed onto it.

\example \figref{state del} demonstrates a type assignment to the user-defined
global state example.

In this type system, the return type of the continuation remains fixed
inside every reset. Existing work on type systems for delimited
control
(\citet{kiselyov-shan:a-substructural-type-system-for-delimited-continuations}
provide a substantial list of references) focuses on type systems that
allow \emph{answer-type modification}, as these can express typed
printf and type-state computation (as in
\citepos*{asai:on-typing-delimited-continuations}{analysis}). We
exclude answer-type modification to keep the fundamental account
clearer and simpler: the type system with answer-type modification is
further removed from the well-known abstractions for effect-handlers
and monadic reflection. We conjecture that the relative expressiveness
of delimited control does not change even with answer-type
modification, once we add analogous capabilities to effect
handlers~\cite{Brady13, kiselyov:parameterized-extensible-effects-and-session-types}
and monadic
reflection~\cite{atkey:parameterised-notions-of-computation}.

Our Abella formalisation establishes:
\begin{theorem}[\lamdel safety]\theoremlabel{del safety}
Well-typed programs don't go wrong: for all closed \lamdel returners
$\Ty\einf\emptyset{}M{\F\Gnd}$, either $M \reducesto N$ for some
$\Ty\einf\emptyset{}{N}{\F\Gnd}$ or else ${M = \return V}$ for some $\Ty\vinf
{}V\Gnd$.
\end{theorem}

Using the translation from $\lamdel$ to $\lammon$ we present in the
next section, $\lamdel$ inherits some of
$\lammon$'s meta-theory.

We define \lamdel's ground types,
plugged contexts and the equivalences $\ceq$ and $\eqcong$ as in
$\cbpv$.

 \newcommand{\transret}[1]{#1^{\mathrm{ret}}}
\newcommand{\transops}[1]{#1^{\mathrm{ops}}}

\section{Macro translations}\label{sec:translations}
\citet{\felleisen} argues that the usual notions of
computability and complexity reduction do not capture the
expressiveness of general-purpose programming languages.
The Church-Turing thesis and its extensions assert that any reasonably
expressive model of computation can be efficiently reduced to any
other reasonably expressive model of computation.  Therefore the
notion of a polynomial-time reduction with a Turing-machine is too
crude to differentiate expressive power of two general-purpose
programming languages.
As an alternative,
Felleisen introduces \emph{macro
  translation}: a \emph{local} reduction of a language extension,
in the sense that it is homomorphic with respect to the
syntactic constructs, and \emph{conservative}, in the sense that it
does not change the core language.
We extend this concept to local translations between conservative
extensions of a shared core.

\paragraph{Translation notation}
We define translations $\macro{\textsc{S}}{\textsc{T}}$ from each
source calculus $\textsc{S}$ to each target calculus $\textsc{T}$.
By default we assume untyped translations, writing $\lameff$,
$\lammon$, and $\lamdel$ in translations that disregard typeability.
In typeability preserving translations (which must also respect the
monad laws where $\lammon$ is concerned) we explicitly write
$\typeable\lameff$, $\proper\lammon$, and $\typeable\lamdel$.
We allow translations to be \emph{hygienic} and introduce fresh
binding occurrences. We write $M \mapsto \trans M$ for the translation
at hand. We include only the non-core cases in the definition
of each translation.

\medskip

Out of the six possible untyped macro-translations, the ideas behind
the following four already appear in the literature:
$\macro{\lamdel}{\lammon}$~\citep{wadler:monads-and-composable-continuations},
$\macro{\lammon}{\lamdel}$~\citep{filinski:representing-monads},
$\macro{\lamdel}{\lameff}$~\citep{bauer-pretnar:programming-with-algebraic-effects-and-handlers},
and $\macro{\lameff}{\lammon}$~\citep{\hia}.
The Abella formalisation contains the proofs of the simulation results
for each of the six translations.
Three translations formally simulate the source calculus by the target
calculus: $\macro\lammon\lamdel$, $\macro\lamdel\lameff$, and
$\macro\lammon\lameff$.
The other translations, $\macro\lamdel\lammon$,
$\macro\lameff\lamdel$, and $\macro\lameff\lammon$, introduce
suspended redexes during reduction that invalidate simulation on the
nose.

For the translations that introduce suspended redexes, we use a
relaxed variant of simulation, namely the relations $\reducestoupto$,
which are the smallest relations containing $\reducesto$ that are
closed under the term formation constructs.  We say that a translation
$M \mapsto \trans M$ is a simulation \emph{up to congruence} if for
every reduction $\M \reducesto \N$ in the source calculus we have
$\trans \M \reducestoupto+\trans \N$ in the target calculus.
In fact, the suspended redexes always $\beta$-reduce by substituting a
variable, i.e., $\force{\thunk{\abst x M}}~x \reducestoupto+\abst x
M$, thus only performing simple rewiring.

\subsection{Delimited continuations as monadic reflection ($\macro{\lamdel}{\lammon}$)}

We adapt Wadler's analysis of delimited
control~\citep{wadler:monads-and-composable-continuations}, using the
continuation monad~\citep{Moggi89}:
\begin{lemma}\lemmalabel{continuation monad}
  For all $\kinf\E\EffectKind$, $\kinf\C{\ComputationKind\E}$, we have
  the following \emph{proper} monad $\Cont$:
  \[
    \kinf{\effType*\E\alpha{\Fun{\U\E{\parent{\Fun\alpha\C}}}\C}
      \effdef*[{\begin{array}[t]{@{}l@{}}}]
          {\x}{\abst
          c{\force c\ x}}{\\m}{f}{\abst c{\force m \
            \thunk{\abst\y{\force f\ \y\ c}}}} }\EffectKind
      \end{array}
  \]
\end{lemma}

Using $\Cont$ we define the macro translation $\macro\lamdel\lammon$
as follows:
\begin{mathpar}
  \trans{\ShiftZ k\M} := \reflect{\abst k {\trans \M}}

  \trans{\Reset \M\x\N} := \reify[\Cont]{\trans{\M}}\ \thunk{\abst \x{\trans \N}}
\end{mathpar}
Shift is interpreted as reflection and reset as reification in
the continuation monad.

\begin{theorem}[$\macro\lamdel\lammon$ correctness]\theoremlabel{DEL->MON correctness}
  $\!\lammon$ simulates $\lamdel\!$ up to congruence: \(
  \M \reducesto \N \implies \trans\M\reducestoupto+\trans\N
  \).
\end{theorem}
The only suspended redex arises in simulating the reflection rule,
where we substitute a continuation into the bind of the continuation
monad yielding a term of the form $\thunk{\abst y{\thunk{\abst
    y M}~y~c}}$ which we must reduce to $\thunk{\abst y{M~c}}$.

$\macro\lamdel\lammon$ extends to a macro
translation at the type level:
\[
\trans{\E,\C} \definedby
\effType
    {\trans\E}
    \alpha
        {\Fun
          {\U
            {\trans\E}
            {\parent{
                \Fun\alpha
                    {\trans\C}}}}
          {\trans\C}}\Cont
\]

\begin{theorem}[$\macro\lamdel\lammon$ preserves typeability]\theoremlabel{del->mon typeability preservation}
  Every well-typed $\lamdel$ phrase $\geninf\Ty\G\E PX$ translates
  into a proper well-typed $\lammon$ phrase: \(
  {\geninf\Ty{\trans\G}{\trans\E}{\trans P}{\trans X}} \).
\end{theorem}

We use this result to extend the meta-theory of $\lamdel$:
\begin{corollary}[$\lamdel$ termination]\corollarylabel{DEL normalisation}
  All well-typed closed ground returners in $\lamdel$ must reduce to a
  unique normal form: if $\cinf[\pure]{}M{\F\Gnd}$ then there exists $V$ such
  that $\vinf {}\V\Gnd$ and \( M \reducesto^\star \return \V
  \).
\end{corollary}

\subsection{Monadic reflection as delimited continuations ($\macro{\lammon}{\lamdel}$)}

We define the macro translation $\macro\lammon\lamdel$ as follows:
\begin{mathpar}
  \trans{\reflect{M}} := \ShiftZ k{\abst b{\force b~(\thunk{\trans{M}},
                                                       \thunk{\abst\x{\force k~x~b}})}}

  \trans{\reify[\effdef*\x{\N_{\mathrm u}}\y f{\N_{\mathrm b}}]{M}}
     := \Reset{\trans{M}}{x}{\abst b{\trans{N_{\textrm u}}}}~\thunk{\abst{\vpair y f}{N_{\textrm b}}}
\end{mathpar}
Reflection is interpreted by capturing the current continuation and
abstracting over the bind operator which is then invoked with the
reflected computation and a function that wraps the continuation in
order to ensure it uses the same bind operator.
Reification is interpreted as an application of a reset. The
continuation of the reset contains the unit of the monad. We apply
this reset to the bind of the monad.

\begin{theorem}[$\macro\lammon\lamdel$ correctness]\theoremlabel{MON->DEL correctness}
  $\lamdel$ simulates $\lammon$ up to congruence: \(
  \M \reducesto \N \implies \trans\M\reducestoupto+\trans\N
  \).
\end{theorem}

This translation does not preserve typeability because the bind
operator can be used at different types. We conjecture that a) any
other macro translation will suffer from the same issue and b) adding
(predicative) polymorphism to the base calculus is sufficient to adapt
this translation to one that does preserve typeability.

\citepos*{filinski:representing-monads}{translation from monadic
  reflection to delimited continuations} does preserve typeability,
but it is a global translation. It is much like our translation except
each instance of bind is inlined (hence it does not need to be
polymorphic).

\subsubsection{Alternative translation with nested delimited continuations}

An alternative to $\macro\lammon\lamdel$ is to use two nested shifts
for reflection and two nested resets for reification:
\begin{mathpar}
  \trans{\reflect{M}} := \ShiftZ k{\ShiftZ b{\force b~(\thunk{\trans{M}},
                                                       \thunk{\abst\x{\Reset{\force k~x}{z}{\force b~z}}})}}

  \trans{\reify[\effdef*\x{\N_{\mathrm u}}\y f{\N_{\mathrm b}}]{M}}
     := \Reset{\Reset{\trans{M}}{x}{\ShiftZ b{\trans{N_{\textrm u}}}}}
                                {\vpair y f}{N_{\textrm b}}
\end{mathpar}
In the translation of reflection, the reset inside the wrapped
continuation ensures that any further reflections in the continuation
are interpreted appropriately: the first shift, which binds $k$, has
popped one continuation off the stack so we need to add one back on.
In the translation of reification, the shift guarding the unit garbage
collects the bind once it is no longer needed.

\subsection{Delimited continuations as effect handlers ($\macro{\lamdel}{\lameff}$)}

We define $\macro\lamdel\lameff$ as follows:
\begin{mathpar}
    \trans{\ShiftZ\k\N} \definedby \op[shift0]\ \thunk{\abst\k{\trans\N}}

    \trans{\Reset\M\x\N} \definedby
        \handle{\trans \M}{{}
    \{\returnClause \x{\trans{\N}}\}
    \uplus{}
    \{\handlerEntry{\op[shift0]}\y f{\force f~y}\}}
\end{mathpar}
Shift is interpreted as an operation and reset is interpreted as a
straightforward handler.

\begin{theorem}[$\macro\lamdel\lameff$ correctness]\theoremlabel{DEL->EFF correctness}
  $\lameff$ simulates $\lamdel$ on the nose: \(
  \M \reducesto \N \implies \trans\M\reducesto+\trans\N
  \).
\end{theorem}

This translation does not preserve typeability because inside a single
reset shifts can be used at different types. We conjecture that a) any
other macro translation will suffer from the same issue and b) adding
polymorphic operations~\cite{\hia} to $\lameff$ is sufficient to
ensure this translation does preserve typeability.

One can adapt our translation to a global translation in which every
static instance of a shift is interpreted as a separate operation,
thus avoiding the need for polymorphic operations.

\subsection{Effect handlers as delimited continuations ($\macro{\lameff}{\lamdel}$)}

We define $\macro\lameff\lamdel$ as follows:
\[
\begin{array}{@{}c@{}}
  \begin{array}{@{}c@{\qquad}c@{}}
  \trans{\op\ V} \definedby
    \ShiftZ k{\abst h{\force h\,(\inj{\op}{\vpair{\trans V}{\thunk{\abst \y{\force k\ \y\ h}}}})}}
  &
  \trans{\handle{M}{H}} \definedby \Resetk{\trans{M}}{\transret{H}}~\thunk{\transops{H}} \\
  \end{array} \bigskip\\

  \begin{array}{@{}c@{\qquad}c@{}}
  \transret{
  \left(
    \begin{array}{@{}l@{}l@{}}
      \begin{array}{@{}r@{~}l@{}}
        \handle {&M}{\\&\{\returnClause\x{\N_{\op[ret]}}\}\,  \\
          \uplus{}& \{\handlerEntry{\op_1}{\p_1}{\k_1}{\N_1}\} \\
          \uplus{}& \ldots \\
          \uplus{}& \{\handlerEntry{\op_n}{\p_n}{\k_n}{\N_n}\}
        }
      \end{array}
    \end{array}
    \right)}
  \!\!\!\!\!\!\definedby
  x.\abst h{\trans{N_{\op[ret]}}}
  &
  \transops{
  \left(
    \begin{array}{@{}l@{}l@{}}
      \begin{array}{@{}r@{~}l@{}}
        \handle {&M}{\\&\{\returnClause\x{\N_{\op[ret]}}\}\,  \\
          \uplus{}& \{\handlerEntry{\op_1}{\p_1}{\k_1}{\N_1}\} \\
          \uplus{}& \ldots \\
          \uplus{}& \{\handlerEntry{\op_n}{\p_n}{\k_n}{\N_n}\}
        }
      \end{array}
    \end{array}
    \right)}
  \!\!\!\!\!\!\definedby
      \begin{array}{@{}r@{}l@{}}
        \abst\y{
          &\Variant{\y}{ \\
            & \VariantCase{\op_1}{\vpair{\p_1}{\k_1}}{\trans{\N_1}}\\
            & \smash\vdots                                        \\
            & \VariantCase{\op_n}{\vpair{\p_n}{\k_n}}{\trans{\N_n}}}
        }
      \end{array}
  \end{array} \\
\end{array}
\]
Operation invocation is interpreted by capturing the current
continuation and abstracting over a dispatcher which is passed an
encoding of the operation. The encoded operation is an injection whose
label is the name of the operation containing a pair of the operation
parameter and a wrapped version of the captured continuation, which
ensures the same dispatcher is threaded through the continuation.

Handling is interpreted as an application of a reset whose
continuation contains the return clause. The reset is applied to a
dispatcher function that encodes the operation clauses.

\begin{theorem}[$\macro\lameff\lamdel$ correctness]\theoremlabel{EFF->DEL correctness}
  $\lamdel$ simulates $\lameff$ up to congruence:
  \(
  \M \reducesto \N  \implies
  \trans\M\reducestoupto+\trans\N
  \).
\end{theorem}
The $\macro\lameff\lamdel$ translation is simpler than Kammar et al.'s
which uses a global higher-order memory cell storing the handler
stack~\citep{\hia}.

This translation does not preserve typeability because the
interpretation of operations needs to be polymorphic in the return
type of the dispatcher over which it abstracts.
We conjecture that a) any other macro translation will suffer from the
same issue and b) adding polymorphism to the base calculus is
sufficient to adapt this translation to one that preserves
typeability.

\subsubsection{Alternative translation with nested delimited continuations}

\newcommand\ED[1]{\mathcal{ED}\!\sem{#1}}

Similarly to the $\macro\lammon\lamdel$ translation there is an
alternative to $\macro\lameff\lamdel$ which uses two nested shifts for
operations and two nested resets for handlers:
\[
\begin{array}{@{}c@{}}
  \begin{array}{@{}c@{\qquad}c@{}}
  \trans{\op\ V} \definedby
    \ShiftZ k{\ShiftZ h{\force h\,(\inj{\op}{\vpair{\trans V}{\thunk{\abst \x{\Reset{\force k\ x}{y}{\force h\ y}}}}})}}
  &
  \trans{\handle{M}{H}} \definedby \Resetk{\Resetk{\trans{M}}{\transret{H}}}{\transops{H}} \\
  \end{array} \bigskip\\

  \begin{array}{@{}c@{\qquad}c@{}}
  \transret{
  \left(
      \begin{array}{@{}r@{}l@{}}
                   &\{\returnClause\x{\N_{\op[ret]}}\}\,  \\
          \uplus{~}& \{\handlerEntry{\op_1}{\p_1}{\k_1}{\N_1}\} \\
          \uplus{~}& \quad\ldots \\
          \uplus{~}& \{\handlerEntry{\op_n}{\p_n}{\k_n}{\N_n}\} \\
      \end{array}
    \right)}
  \!\!\!\!\!\!\definedby
  \x.{\ShiftZ h{\trans{\N_{\op[ret]}}}}
  &
  \transops{
  \left(
      \begin{array}{@{}r@{}l@{}}
                   &\{\returnClause\x{\N_{\op[ret]}}\}\,  \\
          \uplus{~}& \{\handlerEntry{\op_1}{\p_1}{\k_1}{\N_1}\} \\
          \uplus{~}& \quad\ldots \\
          \uplus{~}& \{\handlerEntry{\op_n}{\p_n}{\k_n}{\N_n}\} \\
      \end{array}
    \right)}
  \!\!\!\!\!\!\definedby
      \begin{array}{@{}r@{}l@{}}
          \y.&\Variant{\y}{ \\
             &\VariantCase{\op_1}{\vpair{\p_1}{\k_1}}{\trans{\N_1}}\\
             &\quad\vdots                                        \\
             &\VariantCase{\op_n}{\vpair{\p_n}{\k_n}}{\trans{\N_n}}}
      \end{array} \\
  \end{array} \\
\end{array}
\]

\subsection{Monadic reflection as effect handlers ($\macro{\lammon}{\lameff}$)}

We simulate reflection with an operation and reification with a
handler. Formally, for every anonymous monad $\Monad$ given by
$\effdef*\x{\N_{\mathrm u}}\y f{\N_{\mathrm b}}$ we define
$\macro\lammon\lameff$ as follows:
\begin{mathpar}
    \trans{\reflect\N} \definedby \op[reflect]\ \thunk{\trans\N}

      \trans{\reify\M} \definedby
        \handle{\trans \M}{{}
          \trans\Monad}

  \trans \Monad \definedby
  \begin{aligned}[t]
    \{\returnClause \x{\trans{\N_{\mathrm u}}}\}
    \uplus{}
    \{\handlerEntry{\op[reflect]}\y f{\trans{\N_{\mathrm b}}}\}
  \end{aligned}
\end{mathpar}
Reflection is interpreted as a reflect operation and reification as a
handler with the unit of the monad as a handler and the bind of the
handler as the implementation of the reflect operation.

\begin{theorem}[$\macro\lammon\lameff$ correctness]\theoremlabel{MON->EFF correctness}
  $\lameff$ simulates $\lammon$ on the nose:
  \(
  \M \reducesto \N \implies
  \trans\M \reducesto+ \trans\N
  \).
\end{theorem}

$\macro\lammon\lameff$ does not preserve typeability. For instance,
consider the following computation of type $\F\bit$ using the
environment monad $\Reader$ given on the right:
\[
\begin{array}[t]{@{}l@{\quad}l@{}}
\reify[\Reader]{
  \begin{array}[t]{@{}l@{}}
  \Let b{\reflect{\thunk{\abst {\vpair bf}{b}}}}{\\
  \Let f{\reflect{\thunk{\abst {\vpair bf}{f}}}}{\\
  \force f\ b
    }
  }
}\ \vpair{\inj{\op[true]}\vunit}{\thunk{\abst b{\return b}}}
  \end{array}
&
\kinf[]{
\begin{array}[t]{@{}l@{}}
\effType{\emptyset}{\alpha}{{\vProdType\bit{\U\emptyset{\parent{\bit\to\F\,\bit}}}}\to \F\alpha}{\\\quad
\effdef*[{\begin{array}[t]{@{}l@{}}}]{\x}{\abst e{\return \x}}{\\m}{f}{
  \abst e{\Let \x{\force m\ e}{\force f\ x\ e}}
}
}
}\EffectKind
\end{array}
\end{array}
\end{array}
\]
Its translation into $\lameff$ is not typeable: reflection can
appear at any type, whereas a single operation is monomorphic. We
conjecture that a) this observation can be used to prove that
\emph{no} macro translation $\macro{\proper\lammon}{\typeable\lameff}$
exists and that b) adding polymorphic operations~\cite{\hia} to
$\lameff$ is sufficient for typing this translation.

\subsection{Effect handlers as monadic reflection ($\macro{\lameff}{\lammon}$)}

We define $\macro\lameff\lammon$ as follows:
\[
\begin{array}{@{}c@{}}
  \begin{array}{@{}c@{\qquad}c@{}}
  \trans{\op\ V} \definedby
    \reflect{\abst k{\abst h{\force h\,(\inj{\op}{\vpair{\trans V}{\thunk{\abst \y{\force k\ \y\ h}}}})}}}
  &
  \trans{\handle{M}{H}} \definedby \reify[\Cont]{\trans M}~\thunk{\transret{H}}~\thunk{\transops{H}}
  \end{array} \bigskip\\
  \begin{array}{@{}c@{\qquad}c@{}}
  \left(
    \begin{array}{@{}l@{}}
      \keyword{handle}~M~\keyword{with} \\
      \begin{array}{@{}r@{~}l@{}}
                  & \{\returnClause\x{\N_{\op[ret]}}\}\,  \\
          \uplus{}& \{\handlerEntry{\op_1}{\p_1}{\k_1}{\N_1}\} \\
          \uplus{}& \ldots \\
          \uplus{}& \{\handlerEntry{\op_n}{\p_n}{\k_n}{\N_n}\} \\
      \end{array} \\
    \end{array}
  \right)^{{\mathrm{ret}}}
  \!\!\!\!\!\!\definedby
  \abst x {\abst h{\trans{\N_{\op[ret]}}}}
  &
  \left(
    \begin{array}{@{}l@{}}
      \keyword{handle}~M~\keyword{with} \\
      \begin{array}{@{}r@{~}l@{}}
                  &\{\returnClause\x{\N_{\op[ret]}}\}\,  \\
          \uplus{}& \{\handlerEntry{\op_1}{\p_1}{\k_1}{\N_1}\} \\
          \uplus{}& \ldots \\
          \uplus{}& \{\handlerEntry{\op_n}{\p_n}{\k_n}{\N_n}\} \\
      \end{array} \\
    \end{array}
  \right)^{{\mathrm{ops}}}
  \!\!\!\!\!\!\definedby
      \begin{array}{@{}r@{}l@{}}
        \abst\y{
          &\Variant{\y}{ \\
            & \VariantCase{\op_1}{\vpair{\p_1}{\k_1}}{\trans{\N_1}}\\
            & \smash\vdots                                        \\
            & \VariantCase{\op_n}{\vpair{\p_n}{\k_n}}{\trans{\N_n}}}
        }
      \end{array} \\
  \end{array} \\
  \end{array}
\]
The translation is much like $\macro\lameff\lamdel$, using the
continuation monad in place of first class continuations.

Operation invocation is interpreted by using reflection to capture the
current continuation and abstracting over a dispatcher which is passed
an encoding of the operation. The encoded operation is an injection
whose label is the name of the operation containing a pair of the
operation parameter and a wrapped version of the captured
continuation, which ensures the same dispatcher is threaded through
the continuation.

Handling is interpreted as an application of a reified continuation
monad computation to the return clause and a dispatcher function that
encodes the operation clauses.

\begin{theorem}[$\macro\lameff\lammon$ correctness]\theoremlabel{EFF->MON correctness}
  $\lammon$ simulates $\lameff$ up to congruence:
  \(
  \M \reducesto \N  \implies
  \trans\M\reducestoupto+\trans\N
  \).
\end{theorem}

This translation does not preserve typeability for the same reason as
the $\macro\lameff\lamdel$ translations: the interpretation of
operations needs to be polymorphic in the return type of the
dispatcher over which it abstracts.
We conjecture that a) any other macro translation will suffer from the
same issue and b) adding polymorphism to the base calculus is
sufficient to adapt this translation to one that does preserve
typeability.

\subsubsection{Alternative translation using a free monad}

An alternative to interpreting effect handlers using a continuation
monad is to use a free monad:
\[
\begin{array}{@{}c@{}}
  \begin{array}{@{}c@{\qquad}c@{}}
  \trans{\op\ V} \definedby
    \reflect{\return{(\inj{\op}{\vpair{\trans V}{\abst x{\return x}}})}}
  &
  \trans{\handle{M}{H}}
  \definedby
  H^\star~\reify[H^\dagger]{\trans M}\\
  \end{array} \bigskip\\

  \begin{array}{@{}r@{~}c@{~}l@{}}
  \left(
          \begin{array}{@{}r@{~}l@{}}
                    &\{\returnClause\x{\N_{\op[ret]}}\}  \\
            \uplus{}& \{\handlerEntry{\op_1}{\p_1}{\k_1}{\N_1}\} \\
            \uplus{}& \ldots \\
            \uplus{}& \{\handlerEntry{\op_n}{\p_n}{\k_n}{\N_n}\} \\
          \end{array}
  \right)^\dagger
  &\definedby&
              \begin{array}{@{}l@{}l@{}}
              \keyword{where~} \{
              &\return{x} = \return{(\inj{\mathrm{ret}}{x})}; \\
              &y \bind k =
                \begin{array}[t]{@{}l@{}l@{}}
                \Variant{\y}{
                           &\VariantCase{\textrm{ret}}{x}{\force k~x} \\
                           &\VariantCase{\op_1}{\vpair{\p_1}{k_1}}
                                               {\return{(\inj{\op}{\vpair{p_1}{\abst x{\force{k_1}~x \bind k}}})}}\\
                           &\smash\vdots                                        \\
                           &\VariantCase{\op_n}{\vpair{\p_n}{k_n}}
                                               {\return{(\inj{\op}{\vpair{p_n}{\abst x{\force{k_n}~x \bind k}}})}}}\}
                \end{array} \\
              \end{array} \bigskip\\
  \left(
          \begin{array}{@{}r@{~}l@{}}
                    &\{\returnClause\x{\N_{\op[ret]}}\}  \\
            \uplus{}& \{\handlerEntry{\op_1}{\p_1}{\k_1}{\N_1}\} \\
            \uplus{}& \ldots \\
            \uplus{}& \{\handlerEntry{\op_n}{\p_n}{\k_n}{\N_n}\} \\
          \end{array}
  \right)^\star
  &\definedby&
              \begin{array}{@{}l@{}l@{}}
                h = \abst{y}{
                       \Variant{\y}{
                           &\VariantCase{\textrm{ret}}{x}{\trans{N_{\op[ret]}}} \\
                           &\VariantCase{\op_1}{\vpair{\p_1}{k}}{\Let{k_1}{\return{\thunk{\abst x{\Let{y}{\force{k}~x}{\force{h}~y}}}}}{\trans{N_1}}} \\
                           &\smash\vdots                                        \\
                           &\VariantCase{\op_n}{\vpair{\p_n}{k}}{\Let{k_n}{\return{\thunk{\abst x{\Let{y}{\force{k}~x}{\force{h}~y}}}}}{\trans{N_n}}}}} \\
              \end{array} \\
  \end{array}
\end{array}
\]
Both the bind operation for the free monad $H^\dagger$ and the
function $h$ that interprets the free monad $H^\star$ are recursive.
Given that we are in an untyped setting we can straightforwardly
implement the recursion using a suitable variation of the $Y$
combinator.
This translation does not extend to the typed calculi as they do not
support recursion.
Nevertheless, we conjecture that it can be adapted to a typed
translation if we extend our base calculus to include inductive data
types, as the recursive functions are structurally recursive.

\subsection{Nonexistence results}

\begin{theorem}
  The following macro translations do \emph{not} exist:
  \begin{itemize}
  \item $\macro{\typeable\lameff}{\proper \lammon}$ satisfying: $\M \reducesto \N \implies \trans \M \ceq
    \trans \N$.

  \item $\macro{\typeable\lameff}{\typeable \mathrlap\lamdel\hphantom\lammon}$ satisfying: $\M
    \reducesto \N \implies \trans\M \ceq \trans\N$.
  \end{itemize}
\end{theorem}
Our proof of the first part hinges on the finite denotation
property~(\lemmaref{finite model lammon}). Briefly, assume to the
contrary that there was such a translation. Consider a single effect
operation symbol $\op[tick] : \vUnitType \to \vUnitType$ and the
terms:
\begin{mathpar}
  \op[tick]^0 \definedby \return \vunit

  \op[tick]^{n+1} \definedby {\op[tick] \vunit};{\op[tick]^n}
\end{mathpar}
All these terms have the same type, and by the homomorphic property of
the hypothesised translation, their translations all have the same
type. By the finite denotation property there are two observationally
equivalent translations and by virtue of a macro translation the two
original terms are observationally equivalent in $\lameff$. But every
distinct pair of $\op[tick]^n$ terms is observationally
distinguishable using an appropriate handler. See
\citepos*{forster:thesis}{thesis} for the full details. The second
part follows from \theoremref{del->mon typeability preservation}.

Regarding the remaining four possibilities, we have seen that there is
a typeability-preserving macro translation
$\macro{\typeable\lamdel}{\proper\lammon}$ (\theoremref{del->mon
  typeability preservation}), but we conjecture that there are no
typeability-preserving translations
$\macro{\proper\lammon}{\typeable\lamdel}$,
$\macro{\typeable\lamdel}{\typeable\lameff}$, or,
$\macro{\proper\lammon}{\typeable\lameff}$.

 \section{Conclusion and further work}\label{sec:conclusion}
We have given a uniform family of formal calculi expressing the common
abstractions for user-defined effects: effect handlers (\lameff),
monadic reflection (\lammon), and delimited control (\lamdel) together
with their natural type-and-effect systems. We have used these calculi
to formally analyse the relative expressive power of the abstractions:
monadic reflection and delimited control have equivalent expressivity;
both are equivalent in expressive power to effect handlers when types
are not taken into consideration; and neither abstraction can
macro-express effect handlers and preserve typeability. We have
formalised the more syntactic aspects of our work in the Abella proof
assistant, and have used set-theoretic denotational semantics to
establish inexpressivity results.

Further work abounds. We want to extend each type system until each
translation preserves typeability. We conjecture that adding
polymorphic operations to \lameff would allow it to macro express
\lamdel and \lammon, and that adding polymorphism to \lammon and
\lamdel would allow them to macro express \lameff. We conjecture
polymorphism would also allow \lamdel to macro express \lammon, and
inductive data types with primitive recursion would also allow \lammon
to macro express \lameff.

We are also interested in analysing \emph{global} translations between
these abstractions. In particular, while \lammon and \lamdel allow
reflection/shifts to appear anywhere inside a piece of code, in
practice, library designers define a fixed set of primitives using
reflection/shifts and only expose those primitives to users. This
observation suggests calculi in which each reify/reset is accompanied
by declarations of this fixed set of primitives. We conjecture that
$\lammon$ and $\lamdel$ can be simulated on the nose via a global
translation into the corresponding restricted calculus, and that the
restricted calculi can be macro translated into $\lameff$ while
preserving typeability. Such two-stage translations would give a
deeper reason why so many examples typically used for monadic
reflection and delimited control can be directly recast using effect
handlers. Other global pre-processing may also eliminate
administrative reductions from our translations and establish
simulation on the nose.

We hope the basic type systems we analysed will form a foundation for
systematic further investigation, especially along the following
extensions. Supporting answer-type
modification~\citep{asai:on-typing-delimited-continuations,
  kobori-kameyama-kiselyov-atm-without-tears} can inform more
expressive type system design for effect handlers and monadic
reflection, and account for
type-state~\citep{atkey:parameterised-notions-of-computation} and
session
types~\citep{kiselyov:parameterized-extensible-effects-and-session-types}.
In practice, effect systems are extended with sub-effecting or effect
polymorphism~\citep{
  bauer-pretnar:an-effect-system-for-algebraic-effects-and-handlers,
  pretnar:inferring-algebraic-effects,
  leijen:koka-handlers,
  lindley-hillerstrom:liberating-effects-with-rows-and-handlers,
  lindley-mcbride-mclaughlin:dobedobedo
}.
To these we add effect-forwarding~\citep{\hia} and
rebasing~\citep{\mia}.

We have taken the perspective of a programming language designer
deciding which programming abstraction to select for expressing
user-defined effects. In contrast,
\citet{schrijvers-pirog-wu-jaskelioff} take the perspective of a
library designer for a specific programming language, Haskell, and
compare the abstractions provided by libraries based on monads with
those provided by effect handlers. They argue that both libraries
converge on the same interface for user-defined effects via Haskell's
type-class mechanism.

Relative expressiveness results are subtle, and the potentially
negative results that are hard to establish make them a risky line of
research. We view denotational models as providing a fruitful method
for establishing such inexpressivity results. It would be interesting
to connect our work with that of \citet{
  laird:exceptions-continuations-and-macro-expressiveness
  ,laird:combining-and-relating-control-effects-and-their-semantics
  ,laird2016:combining-control-effects-and-their-models}, who analyses
the macro-expressiveness of a hierarchy of combinations of control
operators and exceptions using game semantics, and in particular uses
such denotational techniques to show certain combinations cannot macro
express other combinations.
We would like to apply similar techniques to compare the expressive
power of local effects such as ML-style reference cells with effect
handlers.

\begin{acks}
Supported by the European Research Council grant `events causality
and symmetry --- the next-generation semantics', and the Engineering
and Physical Sciences Research Council grants EP/N007387/1 `quantum
computing as a programming language' and EP/H005633/1 `Semantic
Foundations for Real-World Systems'.  The material is based upon
work supported by the Air Force Office of Scientific Research, Air
Force Materiel Command, USAF under Award No. FA9550-14-1-0096.
We thank
Andrej Bauer,
Paul Downen,
Marcelo Fiore,
Mathieu Huot,
Kayvan Memarian,
Sean Moss,
Alan Mycroft,
Ian Orton,
Hugo Paquet,
Jean Pichon-Pharabod,
Matthew Pickering,
Reuben Rowe,
Philip Saville,
Sam Staton, and
Tamara von Glehn
for useful suggestions and discussions.
\end{acks}

\bibliographystyle{ACM-Reference-Format}

\end{document}